\documentclass[intlimits,twoside,a4paper]{article}
\usepackage[cp1251]{inputenc}

\usepackage[eqsecnum]{cmpj3}

\usepackage{multirow}
\usepackage{bm}
\DeclareMathOperator{\Sp}{Sp}

\issue{2017}{20}{4}{43001}
\doinumber{10.5488/CMP.20.43001}

\title[Reference system approach]%
{Reference system approach within the white-dwarfs theory}
\author[M.V. Vavrukh, D.V. Dzikovskyi, N.L. Tyshko]{M.V. Vavrukh, D.V. Dzikovskyi, N.L. Tyshko}
\address{%
Department of Astrophysics, Ivan Franko National University of Lviv, \\
8 Kyrylo and Methodiy St., 79005 Lviv, Ukraine
}

\date{Received September 7, 2017, in final form October 5, 2017}

\begin{document}

\maketitle

\begin{abstract}
Reference system approach of non-relativistic electron fluid theory was adapted for calculation of characteristics of the electron-nuclear model at the densities typical of degenerate dwarfs. Two- and three correlation functions of degenerate relativistic electron gas have been calculated in the momentum-frequency representation in the local field approximation.
Main contributions of the Coulomb interactions to the energy and equation of state of the model at $T=0$~K have been calculated in the adiabatic approximation.
\pacs 05.30.Fk, 97.20Rp, 97.60Bw
\end{abstract}

\section{Introduction}
A hundred years have passed since the discovery of degenerate dwarfs \cite{ref01}. The theory of an internal structure of cold dwarfs was developed by  Chandrasekhar in 1940s, and it was based on the equation of state of ideal relativistic electron gas at $T=0$~K \cite{ref02,ref03}. Generalization of this theory followed in the next decades, when in the works of many authors there were investigated effects of such important factors as axial rotation \cite{ref04,ref05}, Coulomb interactions \cite{ref06}, incomplete degeneration of an electron subsystem \cite{ref07,ref08}, effects of magnetic fields \cite{ref09,ref10}, effects of general relativity \cite{ref05,ref11}, processes of neutronization \cite{ref12}, etc. Interpretation of the whole diversity of properties of the dwarfs obtained from the observations of space \cite{ref13,ref14}, requires constructing a general theory that also takes into account the effects of the above mentioned factors, among which there are the competing ones.

The effect of interactions that play an important role in determining the structure of dwarfs at different masses and luminosities, and especially for the case of massive cold dwarfs, is the least studied. Basing on Wigner-Seitz, Thomas-Fermi approximations and non-relativistic random phase approximation corresponding to the approximate accounting of two-particle electron correlations,  Salpeter \cite{ref06} showed that Coulomb interactions lead to a small decrease of pressure of degenerate relativistic electron gas at $T=0$~K, which is still considered to be the basis of  Chandasekhar's theory \cite{ref10}.

Due to the high density ($\sim10^{5}$~g/cm$^{3}$), the matter in the cores of degenerate dwarfs has a metal electron structure with completely collectivized electrons, and the Fermi momentum is of the order $m_{0}c$. Thus,   the non-relativistic approach is not applicable, and it complicates the calculations, because the correlation functions of the reference system (of an interacting relativistic electron gas) are less investigated than the functions of the analogical non-relativistic model. On the other hand, the electron-nuclear model at ``dwarfs'' densities is slightly non-ideal, which allows one to use a random phase approximation in order to account for the interactions. A consistent approach that corresponds to the modern metals many-electron theory is proposed in \cite{ref15}. A disadvantage of this work is the use of approximate expressions for two- and three particle correlation functions of ideal relativistic degenerate electron gas. In the present work we use an accurate static two-electron correlation function and a three-particle correlation function in the long-wavelength approximation, which improves the reliability of the results for the energy of the ground state of the model and equation of state.

\section{General relations}

We consider a more realistic, compared to \cite{ref15},  spatially homogeneous electrically neutral model, which consists of $N_{\text e}$ electrons, $N_{1}$ nuclei of charge  $z_{1}$ and $N_{2}$ nuclei of charge $z_{2}$ in the volume $V$ in thermodynamic limit $N_{\text e},\ V\rightarrow\infty,\ N_{\text e}/V=\text{const}$ at the temperatures much lower than the temperature of the electron subsystem degeneration. Generalization for a larger number of nuclei species is obvious.

For the Hamiltonian of the model
\begin{eqnarray}
\hat{H}=\hat{H}_{0}+\hat{V}_{\text{ee}}+\sum\limits_{i=1}^{2}\hat{V}_{\text{en}}^{i}+\hat{V}_{\text{nn}}\,,
\label{pss01}
\end{eqnarray}
we use a secondary quantization representation for the electrons and a coordinate one for the nuclear subsystem:
\begin{eqnarray}
\hat{H}_{0}=\sum\limits_{\mathbf{k},s}E_{k}\,a_{\mathbf{k},s}^{+}a_{\mathbf{k},s}
\label{pss02}
\end{eqnarray}
is the Hamiltonian of free electrons $(E_{k}=[(m_{0}c^{2})^{2}+\hbar^{2}k^{2}c^{2}]^{1/2}-m_{0}c^{2})$,
\begin{eqnarray}
\hat{V}_{\text{ee}}=\frac{1}{2V}\sum\limits_{\mathbf{q}\ne0}V_{q}\,\hat{I}_{2}(\mathbf{q},-\mathbf{q})
\label{pss03}
\end{eqnarray}
is the operator of electron-electron interactions,
\begin{eqnarray}
\hat{V}_{\text{en}}^{i}=-\frac{1}{V}z_{i}\sum\limits_{\mathbf{q}\ne0}V_{q}\,S_{-\mathbf{q}}^{(i)}\,\hat{\rho}_{\mathbf{q}}
\label{pss04}
\end{eqnarray}
is the operator of electron interactions with $i$-th nuclear subsystem,
\begin{eqnarray}
\hat{V}_{\text{nn}}=\frac{1}{2V}\sum\limits_{\mathbf{q}\ne0}V_{q}\sum\limits_{i,j}z_{i}z_{j}
\big[S_{\mathbf{q}}^{(i)}S_{-\mathbf{q}}^{(j)}-N_{i}\delta_{i,j}\big]
\label{pss05}
\end{eqnarray}
is the sum of direct nuclear interactions. Here, $V_{q}=4\piup e^{2}/q^2$, $S_{\mathbf{q}}^{(i)}=\sum\nolimits_{l=0}^{N_i}\exp{[\ri(\mathbf{q,R}_{l}^{i})]}$ is the structure factor of $i$-th nuclear subsystem,
\begin{align}
\hat{I}_{2}(\mathbf{q},-\mathbf{q})&=\sum\limits_{\mathbf{k}_{1},\mathbf{k}_{2}}\sum\limits_{s_{1},s_{2}}a^{+}_{\mathbf{k}_{1}+\mathbf{q},s_{1}}
a^{+}_{\mathbf{k}_{2}-\mathbf{q},s_{2}}a_{\mathbf{k}_{2},s_{2}}a_{\mathbf{k}_{1},s_{1}},\nonumber\\
\hat{\rho}_{q}&=\sum\limits_{\mathbf{k},s}a^{+}_{\mathbf{k+q},s}a_{\mathbf{k},s}\,,
\label{pss06}
\end{align}
$a^{+}_{\mathbf{k},s},a_{\mathbf{k},s}$ are the creation and  annihilation operators of electrons in quantum states with the given vector~$\mathbf{k}$ and the spin variable $s$, $\mathbf{R}_{l}^{i}$ is the radius-vector of $l$-th nucleus with the charge $z_{i}$.

For the calculation of a partition function of an electron subsystem in the fixed nuclei field in the grand canonical ensemble
\begin{eqnarray}
Z(\mu)=\Sp_{\text e}\exp{\left[-\beta\left(\hat{H}_{0}-\mu \hat{N}_{\text e}+\hat{V}_{\text{ee}}+\sum\limits_{i=1}^{2}\hat{V}_{\text{en}}^{i}\right)\right]},
\label{pss07}
\end{eqnarray}
we summarize the reference system approach developed in  \cite{ref16,ref17} for a description of non-relativistic model of electron liquid. In the formula~\eqref{pss07}, $N_{\text e}$ is the operator of the number of electrons,  $\mu$ is the variable of chemical potential. As in \cite{ref16}, let us move to the ``frequency'' representation of electron operators
\begin{eqnarray}
a_{\mathbf{k},s}(\nu_{*})=\int\limits_{0}^{\beta}a_{\mathbf{k},s}(\beta')\psi_{\nu_{*}}(\beta')\rd\beta',
\label{pss08}
\end{eqnarray}
where $\psi_{\nu_{*}}(\beta')=\beta^{-1/2}\exp{(\ri\nu_{*}\beta')}$, $\nu_{*}=(2n+1)\piup\beta^{-1}$, $n=0;\pm1;\pm2;\ldots.$
In the ``frequency'' representation
\begin{eqnarray}
\exp{\left[-\beta\left(\hat{H}_0-\mu\hat{N}_{\text e}+\hat{V}_{\text{ee}}+\sum\limits_{i}\hat{V}_{\text{en}}^{i}\right)\right]}=
\exp{(-\beta\mathcal{\hat{H}}_{\mu})T_{\beta}\{\hat{S}_{\text{ee}}(\nu)\hat{S}_{\text{en}}(\nu)\}},
\label{pss09}
\end{eqnarray}
and $\mathcal{\hat{H}}_\mu=\hat{H}_{0}-\mu \hat{N}_{\text e}$. The operator $T_{\beta}$ is  a generalized ordering operator,
\begin{align}
\hat{S}_{\text{ee}}(\nu)&=\exp{\bigg[-(2\beta V)^{-1}\sum\limits_{\mathbf{q}\ne0}\sum\limits_{\nu}V_{q}\,\hat{\rho}_{\mathbf{q},\nu}\,\hat{\rho}_{-\mathbf{q},-\nu}\bigg]},\nonumber\\
\hat{S}_{\text{en}}(\nu)&=
\exp{\bigg[V^{-1}\sum\limits_{i=1}^{2}z_{i}\sum\limits_{\mathbf{q}\ne0}V_{q}\,S_{-\mathbf{q}}^{(i)}\,\hat{\rho}_{\mathbf{q},0}\bigg]},\nonumber\\
\hat{\rho}_{\mathbf{q},\nu}&=\sum\limits_{\mathbf{k},s}\sum\limits_{\nu_{*}}a^{+}_{\mathbf{k+q},s}(\nu_{*}+\nu)a_{\mathbf{k},s}(\nu_{*}),
\label{pss10}
\end{align}
moreover, $\nu=2\piup n\beta^{-1}$; $n=0,\pm1,\pm2,\ldots,$ and the calculation of the average of the product of operators $a_{\mathbf{k},s}(\nu_{*})$ is performed according to the rule \cite{ref19,ref20}
\begin{eqnarray}
-\langle T_{\beta}\{a_{\mathbf{k}_{1},s_{1}}(\nu_{*})a^{+}_{\mathbf{k}_{2},s_{2}}(\nu_{*}')\}\rangle_{\mathcal{H}_{\mu}}=
G_{\mathbf{k}_{1},s_{1}}(\nu_{*})\delta_{\nu_{*},\nu_{*}'}\delta_{{s}_{1},s_{2}}\delta_{\mathbf{k}_{1},\mathbf{k}_{2}}\,,
\label{pss11}
\end{eqnarray}
where $G_{\mathbf{k},s}(\nu_{*})=(\ri\nu_{*}-E_{k}+\mu)^{-1}\exp{(\ri\delta\nu_{*})}$ is the spectral image of one-electron Green's function of an ideal system $(\delta\rightarrow+0)$, which is the reference system for a description of the interacting electron gas model.

We use model with the Hamiltonian $H_{0}+\hat{V}_{\text{ee}}$  as a reference system to calculate $Z(\mu)$:
\begin{align}
&Z(\mu)=\exp{[-\beta\Omega_{\text e}(\mu)]}\langle\hat{S}_{\text{en}}(\nu)\rangle_{\text e}\,,\nonumber\\
&\exp{[-\beta\Omega_{\text e}(\mu)]}=\exp{[-\beta\Omega_{0}(\mu)]}\langle T_{\beta}\hat{S}_{\text{ee}}(\nu)\rangle_{0}\,,\nonumber\\
&\langle\hat{S}_{\text{en}}(\nu)\rangle_{\text e}=\langle T_{\beta}\{\hat{S}_{\text{ee}}(\nu)\hat{S}_{\text{en}}(\nu)\}\rangle_{0}[\langle T_{\beta}\hat{S}_{\text{ee}}(\nu)\rangle_{0}]^{-1}.
\label{pss12}
\end{align}
The symbol $\langle\hat{A}\rangle_{0}$ denotes a statistical averaging over the states, and $\langle\hat{A}\rangle_{\text e}$ for the states of the reference system;
\begin{eqnarray}
\Omega_{0}(\mu)=-\beta^{-1}\sum\limits_{\mathbf{k},s}\ln{\{1+\exp{[-\beta(E_{k}-\mu)]}\}}
\label{pss13}
\end{eqnarray}
is the grand potential of an ideal system of electrons, and $\Omega_{\text e}(\mu)$ is the grand potential of the interacting electron gas.

Expanding the operator $\hat{S}_{\text{en}}(\nu)$ in the formula~\eqref{pss12} in a power series, averaging by states for the reference system and presenting the result in an exponential form, we have obtained  contributions to the grand potential of  electron-nuclear interactions
\begin{align}
\Omega_{\text{en}}&=-\sum\limits_{n\geqslant 2}(n!V^{n})^{-1}\sum\limits_{i_1,i_2,\ldots, i_n}z_{i_1}z_{i_2}\ldots z_{i_n}\sum\limits_{\mathbf{q}_{1},\ldots,\mathbf{q}_{n}\ne0}V_{q_{1}}V_{q_{2}}\ldots V_{q_{n}}
S_{-\mathbf{q}_{1}}^{(i_{1})}S_{-\mathbf{q}_{2}}^{(i_{2})}\ldots S_{-\mathbf{q}_{n}}^{(i_{n})}\nonumber\\
&\quad\times\tilde{\mu}_{n}(\mathbf{q}_{1},\mathbf{q}_{2},\ldots, \mathbf{q}_{n}|0)\delta_{\mathbf{q}_{1}+\ldots+\mathbf{q}_{n},0}\,,
\label{pss14}
\end{align}
where
\begin{eqnarray}
\tilde{\mu}_{n}(\mathbf{q}_{1},\ldots, \mathbf{q}_{n}|0)=\beta^{-1}\langle T_{\beta}\{\hat{\rho}_{\mathbf{q}_{1},0}\ \hat{\rho}_{\mathbf{q}_{2},0}\ldots\hat{\rho}_{\mathbf{q}_{n},0}\}\rangle_{\text e}^{\text{c}}
\label{pss15}
\end{eqnarray}
is a cumulant part of the average of the product of $n$ operators $\hat{\rho}_{\mathbf{q},0}$. This is a static limit of $n$-particles correlation function of the reference system
\begin{eqnarray}
\tilde{\mu}_{n}(y_{1},\ldots, y_{n})=\tilde{\mu}_{n}(\mathbf{q}_{1},\ldots, \mathbf{q}_{n}|\nu_{1},\ldots,\nu_{n})
=\beta^{-1}\langle T_{\beta}\{\hat{\rho}_{\mathbf{q}_{1},\nu_{1}}
\hat{\rho}_{\mathbf{q}_{2},\nu_{2}}\ldots\hat{\rho}_{\mathbf{q}_{n},\nu_{n}}\}\rangle_{\text e}^{\text{c}}.
\label{pss16}
\end{eqnarray}
 The functions~\eqref{pss15}, \eqref{pss16} are a generalization of  ``many-tails'' of non-relativistic theory of the electron-ion model of metals \cite{ref16,ref17} and have a well-defined physical meaning. In particular, the functions~\eqref{pss16} are a spectral representation of $n$-particles correlation functions, which are given in the coordinate space. For example, the binary distribution function of the interacting electron gas $F_{2}(\mathbf{r})$ is associated with the function $\tilde{\mu}_{2}(\mathbf{q},-\mathbf{q}|\nu,-\nu)$ of the expression
 \begin{eqnarray}
F_{2}(\mathbf{r})=1+[\beta N_{\text e}(N_{\text e}-1)]^{-1}
\sum\limits_{\nu}\sum\limits_{\mathbf{q}\ne0}
\tilde{\mu}_{2}(\mathbf{q},-\mathbf{q}|\nu,-\nu)\exp{[\ri(\mathbf{q},\mathbf{r})].}
\label{pss17}
\end{eqnarray}

The dynamic correlation functions of $n$-particles \eqref{pss16} are determined by the  polarization operators $M_{n}(y_{1},\ldots,y_{n})$ of $n$-particles \cite{ref17}
\begin{align}
&\mu_{2}(y,-y)=M_{2}(y,-y)\left[1+\frac{V_{q}}{V}M_{2}(y,-y)\right]^{-1},\nonumber\\
&\mu_{n}(y_{1},\ldots, y_{n})=M_{n}(y_{1},\ldots,y_{n})
\prod\limits_{i=1}^{n}\left[1+\frac{V_{q_{i}}}{V}M_{2}(y_{i},-y_{i})\right]^{-1},\quad
n\geqslant 3.
\label{pss18}
\end{align}
These ratios generalize the well-known random phase approximation, in which
$M_{2}(y,-y)\Rightarrow\tilde{\mu}_{2}^{0}(y,-y),\\M_{n}(y_{1},\ldots,y_{n})\Rightarrow\tilde{\mu}_{n}^{0}(y_{1},\ldots,y_{n})$, where
\begin{eqnarray}
\tilde{\mu}_{n}^{0}(y_{1},\ldots, y_{n})=\beta^{-1}\langle T_{\beta}\{\hat{\rho}_{\mathbf{q}_{1},\nu_{1}}\ldots\hat{\rho}_{\mathbf{q}_{n},\nu_{n}}\}\rangle_{0}^{\text{c}}
\label{pss19}
\end{eqnarray}
is the spectral image of correlation functions of an ideal electron gas model,
\begin{align}
&M_{2}(y,-y)=\tilde{\mu}_{2}^{0}(y,-y)\left[1-\frac{V_{q}}{V}\tilde{\mu}_{2}^{0}(y,-y)\,G(y)\right]^{-1},\nonumber\\
&M_{n}(y_{1},\ldots, y_{n})=\tilde{\mu}_{n}^{0}(y_{1},\ldots,y_{n})
\prod\limits_{i=1}^{n}\left[1-\frac{V_{q_{i}}}{V}\tilde{\mu}_{2}^{0}
(y_{i},-y_{i})\,G(y_{i})\right]^{-1}
\label{pss20}
\end{align}
for the $n\geqslant 3$, so the problem of calculating the functions~\eqref{pss16} is reduced to the calculation of correlation functions of the ideal relativistic gas model~\eqref{pss19} and the local field correction function  $G(y)$ for the relativistic interacting electron gas.

\section{Correlation functions of ideal degenerate relativistic electron gas}

Static and dynamic correlation functions of the non-relativistic ideal electron gas are well known. The analytical expression for $\tilde{\mu}_{2}^{0}(y,-y)$ was obtained in \cite{ref18}. The function $\tilde{\mu}_{3}^{0}(\mathbf{q}_{1},\mathbf{q}_{2},-\mathbf{q}_{1}-\mathbf{q}_{2}|0,0,0)$ and the function $\tilde{\mu}_{4}^{0}(\mathbf{q}_{1},-\mathbf{q}_{1},\mathbf{q}_{2},-\mathbf{q}_{2}|0,\ldots,0)$ were calculated in \cite{ref19,ref20,ref21}. The dynamic functions $\tilde{\mu}_{3}^{0}(y_{1},y_{2},-y_{1}-y_{2})$ and  $\tilde{\mu}_{4}^{0}(y_{1},-y_{1},y_{2},-y_{2})$ were first calculated in \cite{ref16}. The calculation of these functions for a relativistic model is a complex problem, because the electron spectrum is not a quadratic function of the wave vector.

According to the definition \eqref{pss19}, the functions $\tilde{\mu}_{n}^{0}(y_{1},\ldots, y_{n})$ are a convolution of one-particle Green's functions, the same as in a non-relativistic case:
\begin{align}
&\mu_{2}^{0}(y,-y)=-\frac{1}{\beta}\sum\limits_{\mathbf{k},s}\sum\limits_{\nu_{*}}G_{\mathbf{k},s}(\nu_{*})G_{\mathbf{k+q},s}(\nu_{*}+\nu),\nonumber\\
&\mu_{3}^{0}(y_{1},y_{2},y_{3})=\frac{2}{\beta}\sum\limits_{\mathbf{k},s}\sum\limits_{\nu_{*}}
G_{\mathbf{k},s}(\nu_{*})G_{\mathbf{k+q}_{1},s}(\nu_{*}+\nu_{1})
G_{\mathbf{k-q}_{2},s}(\nu_{*}-\nu_{2})\delta_{\mathbf{q}_{1}+\mathbf{q}_{2}+
\mathbf{q}_{3},0}\delta_{\nu_{1}+\nu_{2}+\nu_{3},0}\,,\nonumber\\
&\mu_{4}^{0}(y_{1},-y_{1},y_{2},-y_{2})=\frac{1}{\beta}\sum\limits_{\mathbf{k},s}\sum\limits_{\nu_{*}}
G_{\mathbf{k},s}(\nu_{*})G_{\mathbf{k-q}_{1},s}(\nu_{*}-\nu_{1})
\sum_{\sigma=\pm1}G_{\mathbf{k}-\sigma\mathbf{q}_{2},s}(\nu_{*}-\sigma\nu_{2})\nonumber\\
&\qquad\qquad\qquad\qquad\times[2G_{\mathbf{k},s}(\nu_{*})+
G_{\mathbf{k+q}_{1}+\sigma\mathbf{q}_{2},s}(\nu_{*}+\nu_{1}+\sigma\nu_{2})].
\label{pss21}
\end{align}
Factorizing the products of Green's functions and using the ratio
\begin{eqnarray}
\frac{1}{\beta}\sum\limits_{\nu_{*}}G_{\mathbf{k},s}{(\nu_{*})}=n_{\mathbf{k},s}=\{1+\exp{[\beta(E_{k}-\mu)]}\}^{-1},
\label{pss22}
\end{eqnarray}
we obtain representation functions
$\tilde{\mu}_{n}^{0}(y_{1},\ldots, y_{n})$:
\begin{align}
&\mu_{2}^{0}(y,-y)=-2\Re\sum\limits_{\mathbf{k},s}\frac{n_{\mathbf{k},s}}{\ri\nu+E_{\mathbf{k}}-E_{\mathbf{k+q}}}\,,\nonumber\\
&\mu_{3}^{0}(y_{1},y_{2},y_{3})=\delta_{y_{1}+y_{2}+y_{3},0}[\gamma_{3}(y_{1},-y_{2})+
\gamma_{3}(y_{2},-y_{3})+\gamma_{3}(y_{3},-y_{1})],\nonumber\\
&\gamma_{3}(y_{1},y_{2})=2\Re\sum\limits_{\mathbf{k},s}n_{\mathbf{k},s}(\ri\nu_{1}+E_{\mathbf{k}}-E_{\mathbf{k+q}_{1}})^{-1}
(\ri\nu_{2}+E_{\mathbf{k}}-E_{\mathbf{k+q}_{2}})^{-1}\ldots.
\label{pss23}
\end{align}

\subsection{Two-particle correlation function}
\begin{figure}[!b]
\center{\includegraphics[width=0.7\textwidth]{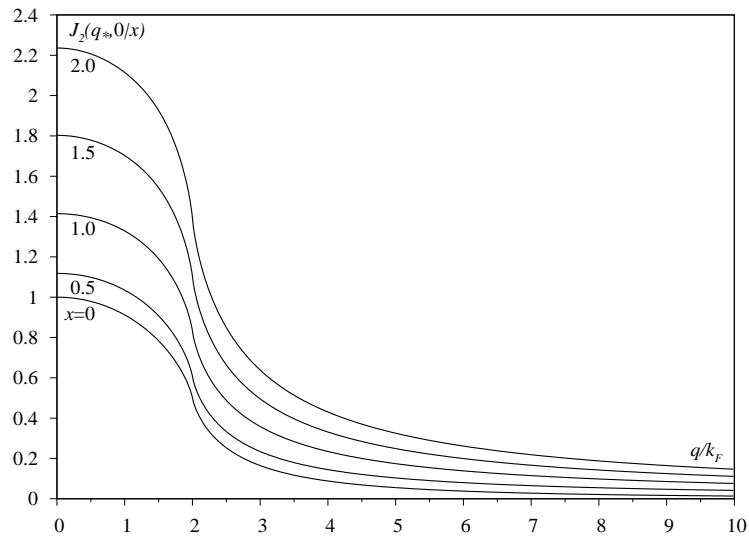}}
\caption{Dependence of the function $J_{2}(q_{*},0|x)$ on the wave vector $\mathbf{q}$ at different values of the relativistic parameter.}
\label{figs01pss}
\end{figure}

 Rewriting the sum over vector $\mathbf{k}$ via integral using a spherical coordinate system and integrating by the angular variables, we obtain a representation:
\begin{align}
\tilde{\mu}_{2}^{0}(y,-y)&=\frac{3N_{\text e}}{m_{0}c^{2}x^{2}}J_{2}(q_{*},\tilde{\nu}|x),\nonumber\\
J_{2}(q_{*},\tilde{\nu}|x)&=\frac{1}{2xq_{*}}\sum\limits_{s}\int\limits_{0}^{\infty}\rd k_{*}k_{*}n_{k_{*},s}A(k_{*}|q_{*},\tilde{\nu}),\nonumber\\
A(k_{*}|q_{*},\tilde{\nu})&=\sum\limits_{\sigma=\pm1}\sigma\biggl\{\big[1+(k_{*}+\sigma q_{*})^{2}\big]^{1/2}-\tilde{\nu}\arctan{\big[\tilde{\nu}\,^{-1}\eta_{\sigma}
(k_{*},q_{*})\big]}\nonumber\\
&\quad
+\frac{1}{2}(1+k_{*}^{2})^{1/2}\ln{[\tilde{\nu}^{2}+\eta_{\sigma}^{2}(k_{*},q_{*})]}\biggr\},\nonumber\\
\eta_{\sigma}(k_{*},q_{*})&=\big[1+(k_{*}+\sigma q_{*})^{2}\big]^{1/2}-(1+k_{*}^{2})^{1/2}.
\label{pss24}
\end{align}
Here, there appear dimensionless variables
\begin{eqnarray}
k_{*}=\frac{x|\mathbf{k}|}{k_{\text F}}\,,\qquad q_{*}=\frac{x|\mathbf{q}|}{k_{\text F}}\,,\qquad \tilde{\nu}=\frac{\nu}{m_{0}c^{2}}\,,
\label{pss25}
\end{eqnarray}
where $x=\hbar k_{\text F}(m_{0}c)^{-1}$ is a relativistic parameter [$k_{\text F}=(3\piup^{2}N_{\text e}/V)^{1/3}$ is the Fermi wave number]. In the static case,

\begin{align}
q_{*}x\,J_{2}(q_{*},0|x)&=\frac{2}{9}(R_{+}-R_{-})\left(1+\frac{7}{4}x^{2}-\frac{q_{*}^{2}}{8}\right)+
\frac{5q_{*}x}{72}(R_{+}+R_{-})+\frac{q_{*}x}{12}R_{0}\nonumber \\
&+\frac{R_{0}^{3}}{3}\ln{\left|\frac{R_{+}-R_{0}}{R_{-}-R_{0}}\right|}+\frac{q_{*}}{8}
\left(1+\frac{q_{*}^{2}}{6}\right)\left[2\ln{|x+R_{0}|}
-\ln{|(R_{+}+x+q_{*})(R_{-}+x-q_{*})|}\right]\nonumber 
\end{align}
\begin{align}
&+\frac{S_{q}^{3}}{6}\Biggl(\ln{\left|\frac{1+\frac{1}{2}q_{*}^{2}+\frac{1}{2}xq_{*}+S_{q}R_{+}}
{1+\frac{1}{2}q_{*}^{2}-\frac{1}{2}xq_{*}+S_{q}R_{-}}\right|}
-\ln{\left|\frac{1+\frac{1}{2}q_{*}x+S_{q}R_{0}}{1-\frac{1}{2}q_{*}x+S_{q}R_{0}}\right|}
-2\ln{\left|\frac{x+\frac{1}{2}q_{*}}{x-\frac{1}{2}q_{*}}\right|}\Biggr),\nonumber\\
R_{0}&=(1+x^{2})^{1/2},\qquad S_{q}=\left(1+\frac{1}{4}q_{*}^{2}\right)^{1/2},\qquad R_{\pm}=\left[1+(q_{*}\pm x)^{2}\right]^{1/2}.
\label{pss26}
\end{align}
As in the non-relativistic case, the function $J_{2}(q_{*},0)$ has a weak logarithmic peculiarity of type $(x-\frac{1}{2}q_{*})\ln{|x-\frac{1}{2}q_{*}|}$ in the vicinity of $q_{*}=2x$ ($|\mathbf{q}|=2k_{\text F}$). In general, the correlation functions of a relativistic model are similar to the corresponding functions of a non-relativistic model, which is clearly visible from the asymptotic of the function $\tilde{\mu}_{2}^{0}(\mathbf{q},-\mathbf{q}|0)$:
\begin{eqnarray}
\tilde{\mu}_{2}^{0}(\mathbf{q},-\mathbf{q}|0)\Rightarrow
\begin{cases}
\dfrac{3N_{\text e}(1+x^{2})^{1/2}}{m_{0}c^{2}x^{2}}+\ldots &\text{by}\,\,q\rightarrow0;\vspace{0.8ex}\\
\dfrac{2N_{\text e}}{m_{0}c^{2}q_{*}}+\ldots&\text{by}\,\,q\rightarrow\infty
\end{cases}
\label{pss27}
\end{eqnarray}
in the relativistic case,
\begin{eqnarray}
\tilde{\mu}_{2}^{0}(\mathbf{q},-\mathbf{q}|0)\Rightarrow
\begin{cases}
\dfrac{3N_{\text e}}{2\varepsilon_{\text F}}+\ldots&\text{by}\,\, q\rightarrow0;\vspace{0.8ex}\\
\dfrac{2N_{\text e}}{\hbar^{2}q^{2}/2m_{0}}+\ldots&\text{by}\,\, q\rightarrow\infty
\end{cases}
\label{pss28}
 \end{eqnarray}
 in the non-relativistic approximation $(2\varepsilon_{\text F}\equiv m_{0}c^{2}x^{2})$. A peculiarity of the function $\tilde{\mu}_{2}^{0}(\mathbf{q},-\mathbf{q}|0)$ is its strong dependence on the relativistic parameter, as shown in figure~\ref{figs01pss}.

In the dynamic case, the $\tilde{\mu}_{2}^{0}(y,-y)$ can be represented as the approximation
\begin{align}
\tilde{\mu}_{2}^{0}(y,-y)&\approx\frac{3N_{\text e}}{m_{0}c^{2}x^{2}}\tilde{J}_{2}(q,v|x),\nonumber \\
\tilde{J}_{2}(q,v|x)&=\frac{1}{4}C(q_{*}|x)\biggl[1-v\left(\arctan\frac{1+q_{*}/2x}{v}
-\arctan\frac{1-q_{*}/2x}{v}\right)\nonumber 
\end{align}
\begin{align}
&\quad+\frac{x}{2q_{*}}\left(1-\frac{q_{*}^{2}}{4x^{2}}+v^{2}\right)\ln\frac{v^{2}+(1+q_{*}/2x)^{2}}{v^{2}+(1-q_{*}/2x)^{2}}
\biggr],\nonumber\\
v&=\nu_{*}(2xq_{*})^{-1}C(q_{*}|x),\qquad C(q_{*}|x)=(1+x^{2}+q_{*}^{2})^{1/2}+(1+x^{2})^{1/2},
\label{pss29}
\end{align}
that only slightly deviates near the maximum from the numerical result, calculated by formula~\eqref{pss24}. In the non-relativistic limit and the long-wavelength approximation $C(q_{*}|x)\rightarrow2$, $v\rightarrow\nu(2\varepsilon_{\text F}q/k_{\text F})^{-1}$, and the function \eqref{pss29} coincides with the known expression for a non-relativistic function $\tilde{\mu}_{2}^{0}(y,-y)$ \cite{ref15}. The expression \eqref{pss29} is here only for completeness. We used the result of numerical calculations for the dynamic function $\tilde{\mu}_{2}^{0}(y,-y)$ to obtain the correlation energy of the reference system (the model of interacting electron gas).

\subsection{Three-particle correlation function}

\begin{figure}[!t]
\center{\includegraphics[width=0.7\textwidth]{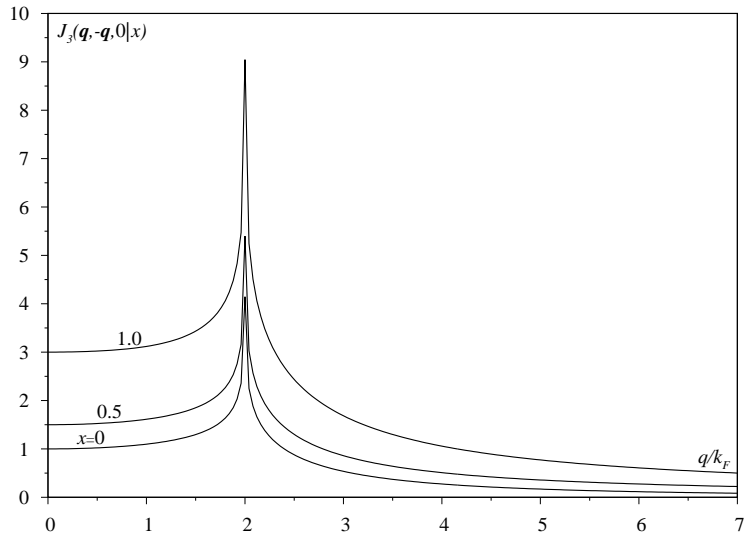}}
\caption{Dependence of the static function $J_{3}(\mathbf{q},-\mathbf{q},0|x)$ on the wave vector $\mathbf{q}$ at different values of the relativistic parameter.}
\label{figs02pss}
\end{figure}

The function $\tilde{\mu}_{3}^{0}(y,-y,0)$, which is a partial case of three-particle dynamic functions when $\mathbf{q}_{2}=-\mathbf{q}_{1}$, $\nu_{2}=-\nu_{1}$, at a full degeneration has an exact analytical image:
\begin{eqnarray}
\tilde{\mu}_{3}^{0}(y,-y,0)=\frac{3N_{\text e}}{(m_{0}c^{2})^{2}q_{*}x^{3}}(1+x^{2})^{1/2}A(x|q_{*},\tilde{\nu}).
\label{pss30}
\end{eqnarray}
In the static limit,
\begin{align}
&\tilde{\mu}_{3}^{0}(\mathbf{q},-\mathbf{q},0|0)=\frac{3N_{\text e}}{(m_{0}c^{2}x^{2})^{2}}J_{3}(\mathbf{q},-\mathbf{q},0|x),\nonumber\\
&J_{3}(\mathbf{q},-\mathbf{q},0|x)=\frac{R_{0}}{\tilde{q}}\left(\tilde{R}_{+}-\tilde{R}_{-}+R_{0}
\ln{\left|\frac{\tilde{R}_{+}-R_{0}}{\tilde{R}_{-}-R_{0}}\right|}\right),\nonumber\\
&\tilde{R}_{\pm}=[1+x^{2}(1\pm\tilde{q})^{2}]^{1/2},\qquad R_{0}=(1+x^{2})^{1/2}.
\label{pss31}
\end{align}
In the formula~\eqref{pss31}, a ``non-relativistic'' scale was used for the wave vector $(\tilde{q}\equiv|\mathbf{q}|/k_{\text F})$. In the long-wavelength limit,
\begin{eqnarray}
\tilde{\mu}_{3}^{0}(\mathbf{q},-\mathbf{q},0|0)=\frac{3N_{\text e}}{(m_{0}c^{2}x^{2})^{2}}(1+2x^{2}).
\label{pss32}
\end{eqnarray}
Dependence of the dimensionless factor $J_{3}(\mathbf{q},-\mathbf{q},0|x)$ on the wave vector and the relativistic parameter is illustrated in figure~\ref{figs02pss}. As in non-relativistic case, the function \eqref{pss31} has a logarithmic peculiarity at $q=2k_{\text F}$.

The formulae~\eqref{pss26}, \eqref{pss29} and \eqref{pss31}, and figures~\ref{figs01pss}, \ref{figs02pss} reveal the general property of correlation functions $\tilde{\mu}_{n}^{0}(y_{1},\ldots,y_{n})$ --- a steep decrease for large wave vectors $(|\mathbf{q}_{i}|>2k_{\text F})$, providing a convergence of integrals in the series \eqref{pss14}.

In general, for a rough estimation of convergence of series \eqref{pss14} we consider a chemically homogeneous model $(z_{1}=z_{2}=z$, $N_{1}+N_{2}=N_{n})$, constraining the integration  for each independent vector $\mathbf{q}_{i}$ of the area $|\mathbf{q}_{i}|<2k_{\text F}$, neglecting the screening interactions, replacing the product of structural factors $S_{\mathbf{q}_{1}}S_{\mathbf{q}_{2}}\ldots S_{\mathbf{q}_{n}}$ with $N_{n}$, we replace the functions $\tilde{\mu}_{n}^{0}(\mathbf{q}_{1},\ldots,\mathbf{q}_{n}|0)$ with $3N_{\text e}(m_{0}c^{2}x^{2})^{1-n}(1+x^{2})^{\frac{1}{2}(n-1)}$, which approximately corresponds to the long-wavelength asymptotic. For the magnitude of $n$-th member of series \eqref{pss14} it can be estimated as
\begin{eqnarray}
N_{\text e}m_{0}c^{2}z^{n-1}\alpha_{0}^{n}x^{2-n}(1+x^{2})^{\frac{1}{2}(n-1)},
\label{pss33}
\end{eqnarray}
where $\alpha_{0}=e^{2}/\hbar c$ is the fine structure parameter. Hence, the series \eqref{pss14} is an expansion for a dimensionless parameter $z\alpha_{0}$, which varies from 0{.}014 (helium dwarf) to 0{.}19 (iron dwarf). For the typical dwarfs, mainly consisting of nitrogen and oxygen, $z\alpha_{0}\approx0{.}1$. That expansion parameter is a small value, which makes it possible to restrict the consideration to two- and three-electron correlations (we note, that correlation energy of the reference system is of the order of $\alpha_{0}^{2}$). Moreover, for the three-electron function $\tilde{\mu}_{3}^{0}(\mathbf{q}_{1},\mathbf{q}_{2},-\mathbf{q}_{1}-\mathbf{q}_{2}|0)$, there can be used an approximate analytical expression, because the main contributions provide the two-electron correlations, and the contribution of  three-electron correlations play a role of a correction.

The calculation of correlation functions $\tilde{\mu}_{3}^{0}(y_{1},y_{2},-y_{1}-y_{2})$ and $\tilde{\mu}_{4}^{0}(y_{1},-y_{1},y_{2},-y_{2})$ in the static or dynamic cases in the non-relativistic theory is based on Feynman identity \cite{ref23}, which allows one to integrate over angular variables of vector $\mathbf{k}$ at a fixed configuration of vectors $\mathbf{q}_{1}$ and $\mathbf{q}_{2}$ in  terms of $\gamma_{3}(y_{i},y_{j})$. Unfortunately, this identity cannot be used  in the exact calculation due to the complex dependence of relativistic electron energy on the wave vector.

In order to make an approximate  calculation using the identity transformation, we present $\gamma_{3}(\mathbf{q}_{i},\mathbf{q}_{j})$ as
\begin{eqnarray}
\gamma_{3}(\mathbf{q}_{i},\mathbf{q}_{j})=2\sum\limits_{\mathbf{k},s}n_{\mathbf{k},s}(\tilde{E}_{\mathbf{k}}+\tilde{E}_{\mathbf{k+q}_{i}})
(\tilde{E}_{\mathbf{k}}+\tilde{E}_{\mathbf{k+q}_{j}})
(\hbar c)^{-4}[2(\mathbf{k},\mathbf{q}_{i})+q_{i}^{2}]^{-1}[2(\mathbf{k},\mathbf{q}_{j})+q_{j}^{2}]^{-1},
\label{pss34}
\end{eqnarray}
where $\tilde{E}_{k}=[(m_{0}c^{2})^{2}+\hbar^{2}c^{2}k^{2}]^{1/2}$. Then, we use the approximation
\begin{eqnarray}
\tilde{E}_{k}+\tilde{E}_{\mathbf{k+q}_{i}}\Rightarrow m_{0}c^{2}C(\tilde{q}_{i}|\tilde{k}),\qquad
C(\tilde{q}_{i}|\tilde{k})=[1+x^{2}(\tilde{k}^{2}+\tilde{q}_{i}^{2})]^{1/2}+(1+x^{2}\tilde{k}^{2})^{1/2},
\label{pss35}
\end{eqnarray}
which is asymptotically correct both at small and at large $q_{i}$. According to the Feynman identity,
\begin{align}
&\frac{1}{[2(\mathbf{k},\mathbf{q}_{i})+q_{i}^{2}][2(\mathbf{k},\mathbf{q}_{j})+q_{j}^{2}]}=
\int\limits_{0}^{1}\frac{\rd z}{F^{2}(\mathbf{q}_{i},\mathbf{q}_{j}|\mathbf{k})}\,,\nonumber\\
&F(\mathbf{q}_{i},\mathbf{q}_{j}|\mathbf{k})=
\Omega_{ij}+2(\mathbf{k},\boldsymbol{\rho}_{ij}),\qquad
\Omega_{ij}\equiv q_{j}^{2}+z(q_{i}^{2}-q_{j}^{2});\qquad\boldsymbol{\rho}_{ij}=z\mathbf{q}_{i}+(1-z)\mathbf{q}_{j}\,.
\label{pss36}
\end{align}
 Rewriting the sum over vector $\mathbf{k}$ via integral, we use dimensionless variable $\tilde{k}=|\mathbf{k}|/k_{\text F},\,\,\tilde{q}_{i}=|\mathbf{q}_{i}|/k_{\text F}$, and the spherical coordinate system, the $Oz$ axis of which coincides with the vector $\boldsymbol{\rho}_{ij}$, we perform integration over the angular variables, reducing $\gamma_{3}(\mathbf{q}_{i},\mathbf{q}_{j})$ to one-dimensional integral:
 \begin{align}
&\gamma_{3}(\mathbf{q}_{i},\mathbf{q}_{j})=\frac{3N_{\text e}}{4(m_{0}c^{2}x^{2})^{2}}\int\limits_{0}^{1}\rd k\,C(q_{i}|k)\,C(q_{j}|k)f_{ij}(k), \nonumber 
\end{align}
\begin{align}
&f_{ij}(k)=\frac{1}{\sqrt{-\delta(k)}}\ln\left|\frac{R_{ij}+[-\delta(k)]^{-1/2}}{R_{ij}-[-\delta(k)]^{-1/2}}\right|\quad \text{by}\,\, k<q_{\text R};\nonumber\\
&f_{ij}(k)=\frac{2}{\sqrt{\delta(k)}}\arctan[\delta^{1/2}(k)R^{-1}_{ij}]\quad \text{by}\,\, k>q_{\text R}.
\label{pss37}
\end{align}
Here, the following notations are introduced:
\begin{align}
R_{ij}\equiv R_{ij}(k)=2(\mathbf{q}_{i},\mathbf{q}_{j})-\frac{q_{i}^{2}q_{j}^{2}}{2k^{2}}&;\qquad
\delta(k)=\delta_{ij}(k)=\left(1-\frac{q_{\text R}^{2}}{k^{2}}\right)4q_{i}^{2}q_{j}^{2}(1-t_{ij}^{2});\nonumber\\
q_{\text R}&=(\mathbf{q}_{i}-\mathbf{q}_{j})^{2}[4(1-t_{ij}^{2})]^{-1};
\label{pss38}
\end{align}
$\delta(k)$ is the invariant of the problem $[\delta_{12}(k)=\delta_{23}(k)=\delta_{31}(k)]$, $q_{\text R}$ is the radius of the circle, circumscribing the triangle constructed on the vectors $\mathbf{q}_{1}, \mathbf{q}_{2},-\mathbf{q}_{1}-\mathbf{q}_{2}$; $t_{ij}$ is a cosine of the angle between the vectors $\mathbf{q}_{i}, \mathbf{q}_{j}$. In the formulae~\eqref{pss37}, \eqref{pss38}, the variables $k$ and $q_{i}, q_{j}$ are dimensionless (in unit $k_{\text F}$).

\begin{figure}[!t]
\center{\includegraphics [width=0.7\textwidth]{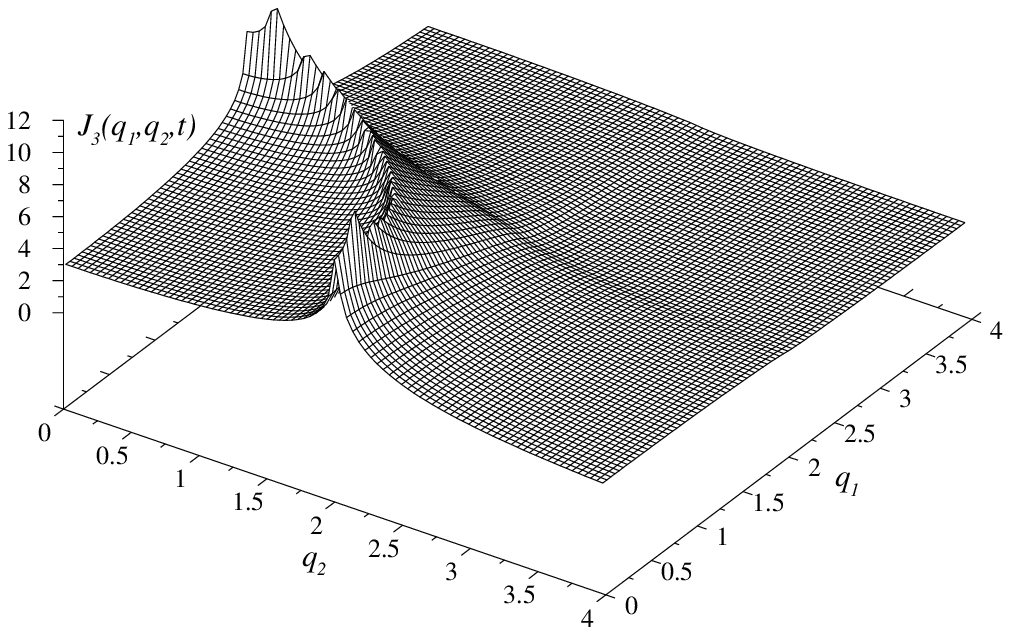}}
\caption{The function $J_{3}({q}_{1},{q}_{2},t)=\tilde{\mu}_{3}^{0}(\mathbf{q}_{1},\mathbf{q}_{2},-\mathbf{q}_{1}-\mathbf{q}_{2})
(3N_{\text e})^{-1}(m_{0}c^{2}x^{2})^{2}$ at $t=0$ and the relativistic parameter $x=1.0$.}
\label{figs03pss}
\end{figure}

\begin{figure}[!t]
\center{\includegraphics [width=0.7\textwidth]{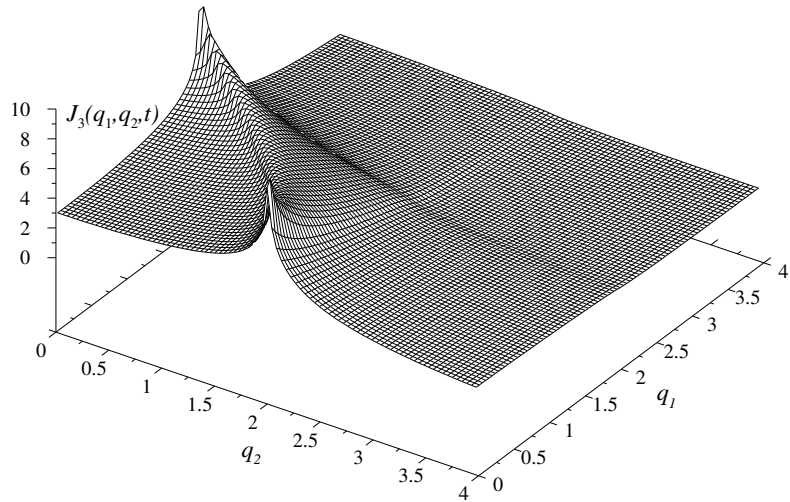}}
\caption{The function $J_{3}({q}_{1},{q}_{2},t)=\tilde{\mu}_{3}^{0}(\mathbf{q}_{1},\mathbf{q}_{2},-\mathbf{q}_{1}-\mathbf{q}_{2})
(3N_{\text e})^{-1}(m_{0}c^{2}x^{2})^{2}$ at $t=0.5$ and the relativistic parameter $x=1.0$.}
\label{figs04pss}
\end{figure}

The substitution $C(q_{i}|k)\rightarrow2(1+x^{2}k^{2})^{1/2}$ corresponds to long-wavelength approximation, which allows one to rewrite $\tilde{\mu}_{3}^{0}(\mathbf{q}_{1},\mathbf{q}_{2},\mathbf{q}_{3}|0)$ in a compact form:
\begin{align}
&\tilde{\mu}_{3}^{0}(\mathbf{q}_{1},\mathbf{q}_{2},\mathbf{q}_{3}|0)\cong\frac{3N_{\text e}}{(m_{0}c^{2}x^{2})^{2}}\int\limits_{0}^{1}
\rd k(1+x^{2}k^{2})\Phi(k|{q_{1},q_{2},q_{3}}),\nonumber\\
&\Phi(k|{q_{1},q_{2},q_{3}})=\frac{k}{a_{1}(k)}\frac{1}{q_{1}q_{2}q_{3}}\ln\biggl|
\frac{1+a_{1}(k)D(k)}{1-a_{1}(k)D(k)}\biggr|\quad\text{by}\,\, k<q_{\text R};\nonumber\\
& \Phi(k|{q_{1},q_{2},q_{3}})=\frac{2k}{a_{2}(k)}\frac{1}{q_{1}q_{2}q_{3}}\arctan[a_{2}(k)D(k)]\quad
\text{by}\,\, k>q_{\text R}.
\label{pss39}
\end{align}
In this formula,
\begin{align}
&a_{1}(k)=\left(1-\frac{k^{2}}{q_{\text R}^{2}}\right)^{1/2};\qquad a_{2}(k)=\left(\frac{k^{2}}{q_{\text R}^{2}}-1\right)^{1/2};\nonumber\\
&D(k)=\frac{q_{1}q_{2}q_{3}}{4k^{3}}\left[1-\frac{(q_{1}^{2}+q_{2}^{2}+q_{3}^{2})}{8k^{2}}\right]\frac{1}{P(k)};\nonumber\\
&P(k)=1-\frac{q_{1}^{2}+q_{2}^{2}+q_{3}^{2}}{4k^{2}}+\frac{q_{1}^{4}+q_{2}^{4}+q_{3}^{4}}{32k^{4}}+\frac{(q_{1}q_{2}q_{3})^{2}}{64k^{6}}.
\label{pss40}
\end{align}
Dependence of function $\tilde{\mu}_{3}^{0}(\mathbf{q}_{1},\mathbf{q}_{2},-\mathbf{q}_{1}-\mathbf{q}_{2}|0) (3N_{\text e})^{-1}(m_{0}c^{2}x^{2})^{2}$ on wave vectors for different $t$ are illustrated in figures~\ref{figs03pss}, \ref{figs04pss}.


\subsection{The local field correction function}

\begin{figure}[!t]
\center{\includegraphics[width=0.7\textwidth]{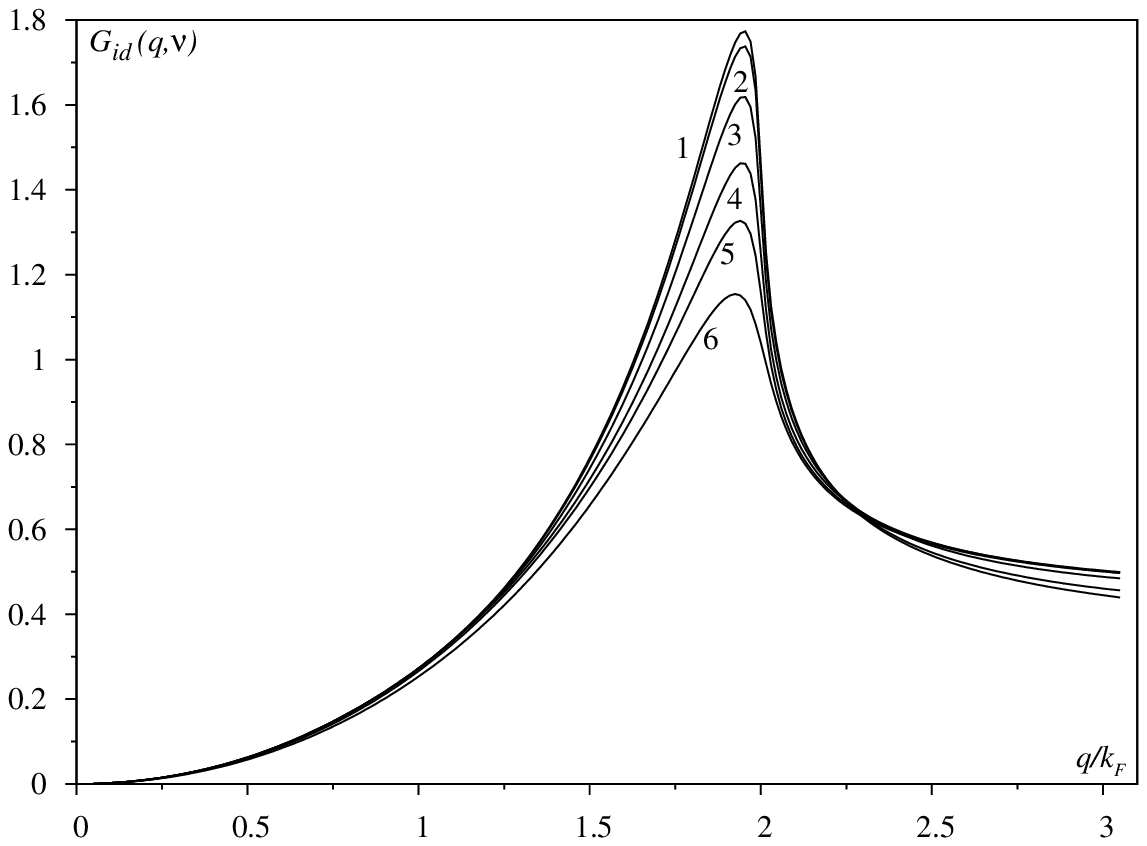}}
\caption{Dependence of the LFCF $G_{\text{id}}(q,\nu)$ on the wave vector $\mathbf{q}$ at frequency $\nu=0.01m_{0}c^{2}x^{2}$ (curve~1~--- $x=0.05$; 2~---~$x=0.2$; 3~---~$x=0.5$; 4~---~$x=1.0$; 5~---~$x=2.0$; 6~---~$x=5.0$).}
\label{figs05pss}
\end{figure}

\begin{figure}[!t]
\center{\includegraphics[width=0.7\textwidth]{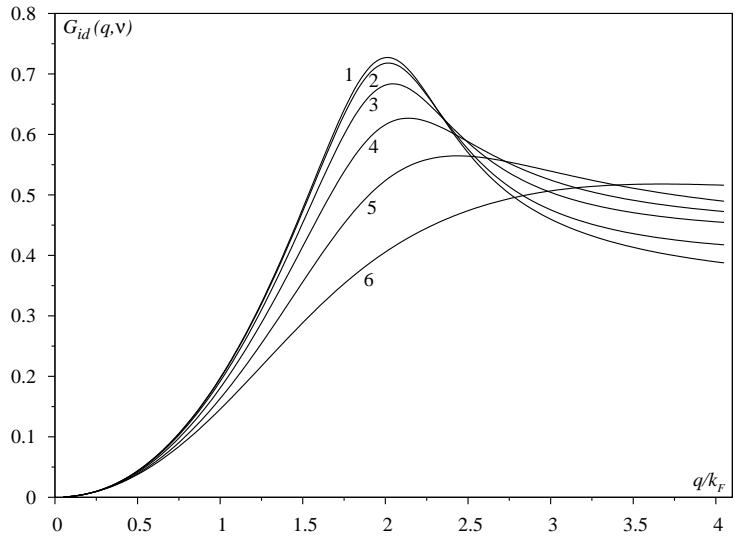}}
\caption{Dependence of the LFCF $G_{\text{id}}(q,\nu)$ on the wave vector $\mathbf{q}$ at frequency $\nu=0.5m_{0}c^{2}x^{2}$ (curve~1~---~$x=0.05$; 2~---~$x=0.2$; 3~---~$x=0.5$; 4~---~$x=1.0$; 5~---~$x=2.0$; 6~---~$x=5.0$).}
\label{figs06pss}
\end{figure}

 It is well known from the non-relativistic electron fluid theory \cite{ref16} that the local field correction function (LFCF) in the weak non-ideal  model is a universal function of the variable $y=(\mathbf{q},\nu)$. It does not depend on any parameters and corresponds to the approximation
\begin{eqnarray}
G_{\text{id}}(y)=-\frac{\sum\nolimits_{y_{1}}V_{q_{1}}\tilde{\mu}_{4}^{0}(y,-y, y_{1},-y_{1})}{2\beta V_{q}[\tilde{\mu}_{2}^{0}(y,-y)]^{2}}\,.
\label{pss41}
\end{eqnarray}
We have calculated the LFCF, using the correlation functions of the relativistic ideal electron model. Summarizing  over the frequencies $\nu_{1}$ and $\nu_{*}$ (which appears in $\tilde{\mu}_{4}^{0}(y,-y, y_{1},-y_{1}$) and using the formula~\eqref{pss22}, we obtain a representation:
\begin{align}
&G_{\text{id}}(y)=-V_{q}^{-1}[\tilde{\mu}_{2}^{0}(y,-y)]^{-2}\sum\limits_{s}\sum\limits_{\mathbf{k}_{1},\mathbf{k}_{2}}
n_{\mathbf{k}_{1},s}n_{\mathbf{k}_{2},s}
\left[V(\mathbf{k}_{1}-\mathbf{k}_{2})f^{-}_{{\mathbf{k}_{1},\mathbf{k}_{2}}}(\mathbf{q},\nu)
-V(\mathbf{k}_{1}+\mathbf{k}_{2}+
\mathbf{q})f^{+}_{{\mathbf{k}_{1},\mathbf{k}_{2}}}(\mathbf{q},\nu)\right],\nonumber\\
&f^{\mp}_{{\mathbf{k}_{1},\mathbf{k}_{2}}}(\mathbf{q},\nu)=\Re\left[(\ri\nu+E_{\mathbf{k}_{1}}-E_{\mathbf{k}_{1}+\mathbf{q}})^{-1}
\mp(\pm \ri\nu+E_{\mathbf{k}_{2}}-E_{\mathbf{k}_{2}+\mathbf{q}})^{-1}\right]^{2}.
\label{pss42}
\end{align}
The behavior of relativistic LFCF at small and large values of vector $\mathbf{q}$ is the same as that of non-relativistic case \cite{ref22}:
\begin{eqnarray}
G_{\text{id}}(y)\Rightarrow
\begin{cases}
\dfrac{(q/k_{\text F})^{2}}{4}+\ldots \quad \text{by}\,\,\nu=0;\,\, q\ll k_{\text F};\\
\dfrac{3(q/k_{\text F})^{2}}{20}+\ldots \quad \text{by}\,\,\nu\gg\varepsilon_{\text F};\,\, q\ll k_{\text F};\\
\dfrac{1}{3}+\ldots\quad\text{by}\,\, q\gg k_{\text F}\,\,\text{and \,any}\,\,\nu.
\end{cases}
\label{pss43}
\end{eqnarray}
For the numerical calculation at the absolute zero temperature, we have used a cylindrical coordinate
system for the vectors $\mathbf{k}_{1},\mathbf{k}_{2}$ [$\mathbf{k}_{j}=(z_{j},\rho_{j},\varphi_{j})$, $k_{j}^{2}=z_{j}^{2}+\rho_{j}^{2}$], in which $(\mathbf{k}_{j},\mathbf{q})=qz_{j}$, $(\mathbf{k}_{1}-\mathbf{k}_{2})^{2}=\rho_{1}^{2}+\rho_{2}^{2}+(z_{1}-z_{2})^{2}-2\rho_{1}\rho_{2}\cos{(\varphi_{1}-\varphi_{2})}$, $(\mathbf{k}_{1},\mathbf{k}_{2})=z_{1}z_{2}+\rho_{1}\rho_{2}\cos{(\varphi_{1}-\varphi_{2})}$. Integrating over the angular variables $\varphi_{1},\varphi_{2}$, we reduce $G_{\text{id}}(y)$ to the following 4-dimensional integral:
\begin{align}
G_{\text{id}}(q,\nu)&=\frac{q^{2}x^{4}}{8}J_{2}^{-2}(qx,\tilde{\nu} x^{2}|x)\int\limits_{-1}^{1}\rd z_{1}\int\limits_{-1}^{1}\rd z_{2}
\int\limits_{0}^{\sqrt{1-z_{1}^{2}}}\rho_{1}\rd\rho_{1}\int\limits_{0}^{\sqrt{1-z_{2}^{2}}}\rho_{2}\rd\rho_{2}\nonumber\\
&\quad\times\left[\frac{f^{+}_{q,\nu}(z_{1},z_{2},\rho_{1},\rho_{2})}{W_{+}(z_{1},z_{2},\rho_{1},\rho_{2})}-
\frac{f^{-}_{q,\nu}(z_{1},z_{2},\rho_{1},\rho_{2})}{W_{-}(z_{1},z_{2},\rho_{1},\rho_{2})}\right].
\label{pss44}
\end{align}
Here, the following notations are used:
\begin{align}
&f^{\pm}_{q,\nu}(z_{1},z_{2},\rho_{1},\rho_{2})=\left(\frac{\eta_{1}}{\eta_{1}^{2}+\bar{\nu}^{2}}\pm
\frac{\eta_{2}}{\eta_{2}^{2}+\bar{\nu}^{2}}\right)^{2}
-\bar{\nu}^{2}\left(\frac{1}{\eta_{1}^{2}+\bar{\nu}^{2}}-
\frac{1}{\eta_{2}^{2}+\bar{\nu}^{2}}\right)^{2},\nonumber\\
&\eta_{i}=[1+x^{2}(z_{i}^{2}+\rho_{i}^{2})]^{1/2}-\left\{1+x^{2}[\rho_{i}^{2}+(z_{i}+q)^{2}]\right\}^{1/2},\nonumber\\
&W_{+}(z_{1},z_{2},\rho_{1},\rho_{2})=\left[(z_{1}+z_{2}+q)^{4}+(\rho_{1}^{2}-\rho_{2}^{2})^{2}+2(\rho_{1}^{2}+
\rho_{2}^{2})(z_{1}+z_{2}+q)^{2}\right]^{1/2},\nonumber\\
&W_{-}(z_{1},z_{2},\rho_{1},\rho_{2})=\left[(z_{1}-z_{2})^{4}+(\rho_{1}^{2}-\rho_{2}^{2})^{2}
+2(\rho_{1}^{2}+\rho_{2}^{2})(z_{1}-z_{2})^{2}\right]^{1/2}.
\label{pss45}
\end{align}
For a comparison $G_{\text{id}}(q,\nu)$ with LFCF of non-relativistic theory, there were used the variables $q=|\mathbf{q}|k_{\text F}^{-1}$, $\bar{\nu}=x^{2}\tilde{\nu}$, $\tilde{\nu}=\nu(\hbar^{2}k_{\text F}^{2}/m)^{-1}$. In figure~\ref{figs05pss} there is shown a function $G_{\text{id}}(q,\nu)$, calculated by the formula~\eqref{pss44} at $\tilde{\nu}=0{.}01$, which is very close to the static limit.
It is obvious that the asymptotic behavior $G_{\text{id}}(q,\nu)$ for small and large values of the $|\mathbf{q}|$ almost does not depend on the relativistic parameter. The deviation of a relativistic correction from non-relativistic one is significant near its maximum, which monotonously decreases with an increase of relativistic parameter. Figure~\ref{figs06pss} illustrates the behavior of $G_{\text{id}}(q,\nu)$ at a very high value of frequency $(\tilde{\nu}=0{.}5)$. In general, the behavior of the correction corresponds to the one of non-relativistic theory, and some deviations are caused by frequency renormalization at $x\geqslant 1$.


\section{The energy of ground state}

As we know from the theory of non-relativistic electron fluid, the transition from thermodynamic description to the quantum mechanics is performed by the following procedure: instead of the chemical potential of interacting system $\mu$ there should be used a chemical potential of the ideal system $\mu_{0}$, and in the $\Omega(\mu)$ there should be considered only the so-called ``normal'' components, because the chemical potential shift $\mu-\mu_{0}$ is compensated with contributions of the so-called ``abnormal'' components, which arise in the correlation functions $\tilde{\mu}_{n}^{0}(y_{1},\ldots, y_{n})$ at $n\geqslant 4$. In the Green's function approach, this fact was first discovered in \cite{ref24} and in the reference system approach --- in \cite{ref17} within the perturbation theory.

\begin{figure}[!b]
\center{\includegraphics[width=0.7\textwidth]{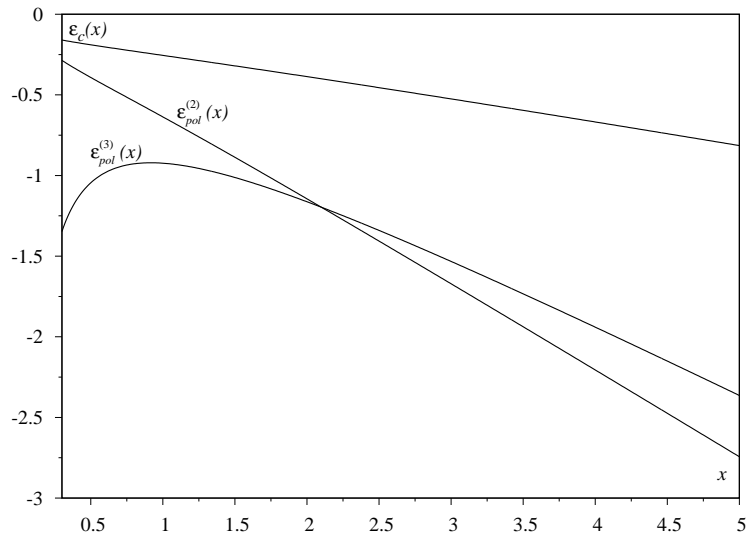}}
\caption{Dependence of functions $\varepsilon^{(2)}_{\text{pol}}(x)$, $10\cdot\varepsilon^{(3)}_{\text{pol}}(x)$ and $\varepsilon_{\text c}(x)$ on the relativistic parameter.}
\label{figs07pss}
\end{figure}

Taking into account a weak non-ideal matter of dwarfs, we will consider only two- and three-particle electron correlations. In this approach, the ground state energy of the model is given as follows:
\begin{align}
E&\simeq E_{\text e}+\frac{1}{2V}\sum\limits_{\mathbf{q}\ne0}V_{q}\sum\limits_{i,j}z_{i}z_{j}\left[S^{(i)}_{\mathbf{q}}S^{(j)}_{-\mathbf{q}}-
N_{i}\delta_{i,j}\right]
-\sum\limits_{n=2}^{3}(n!V^{n})^{-1}\sum\limits_{i_{1},i_{2},\ldots,i_{n}}z_{i_{1}}z_{i_{2}}\ldots z_{i_{n}}\nonumber\\
&\quad\times\sum\limits_{\mathbf{q}_{1},\mathbf{q}_{1},\ldots, \mathbf{q}_{n}\ne0}
 V_{q_{1}}\ldots V_{q_{n}}S^{(i_{1})}_{-\mathbf{q}_{1}}\ldots S^{(i_{n})}_{-\mathbf{q}_{n}}
\delta_{\mathbf{q}_{1}+\ldots+\mathbf{q}_{n},0}\,\tilde{\mu}_{n}(\mathbf{q}_{1}\ldots\mathbf{q}_{n}|0).
\label{pss46}
\end{align}
Here, $E_{\text e}$ is the ground state energy of the reference system (an interacting relativistic electron gas), which can be calculated using the expression
\begin{eqnarray}
E_{\text e}=E_{0}+\frac{1}{2\beta V}\sum\limits_{\mathbf{q}\ne0}\sum\limits_{\nu}V_{q}\int\limits_{0}^{1}\tilde{\mu}_{2}^{\lambda}(y,-y)\rd\lambda,
\label{pss47}
\end{eqnarray}
where $\tilde{\mu}_{2}^{\lambda}(y,-y)$ is the two-particle dynamic correlation function of the auxiliary model of electrons with the potential of interactions $\lambda V_{q}$, and $E_{0}$ is the energy of the ideal electron system
\begin{eqnarray}
E_{0}=\sum\limits_{\mathbf{k},s}n_{\mathbf{k},s}(\mu_{0})E_{k}.
\label{pss48}
\end{eqnarray}
Extracting from the second term of  formula~\eqref{pss47} the contribution of ideal correlation
\begin{eqnarray}
E_{\text{HF}}=-\frac{1}{2\beta V}\sum\limits_{\mathbf{q}\ne0}V_{q}\sum\limits_{\nu}\tilde{\mu}_{2}^{0}(y,-y)
=-\frac{1}{2V}\sum\limits_{\mathbf{q}\ne0}\sum\limits_{\mathbf{k},s}V_{q}\,n_{\mathbf{k}+\mathbf{q}/2,s}n_{\mathbf{k}-\mathbf{q}/2,s}
\label{pss49}
\end{eqnarray}
and considering, that $G_{\text{id}}(y)$ does not depend on the ``coupling constant'' $\lambda$, we have represented $E_{\text e}$ in a traditional form
\begin{eqnarray}
E_{\text e}=E_{0}+E_{\text{HF}}+E_{\text c}\,,
\label{pss50}
\end{eqnarray}
where
\begin{eqnarray}
E_{\text c}=\frac{1}{2\beta}\sum\limits_{\mathbf{q}\ne0}\sum\limits_{\nu}\frac{\ln[1+L(y)]-L(y)}{1-G_{\text{id}}(y)}\,,\qquad
L(y)=\frac{V_{q}}{V}\tilde{\mu}_{2}^{0}(y,-y)[1-G_{\text{id}}(y)]
\label{pss51}
\end{eqnarray}
is the so-called correlation energy (the contribution of non-ideal correlations). In units $m_{0}c^{2}$ we obtain
\begin{eqnarray}
E_{\text e}=N_{\text e}m_{0}c^{2}\left[\varepsilon_{0}(x)-\frac{3}{4\piup}\alpha_{0}x+\alpha_{0}^{2}\,\varepsilon_{\text c}(x)\right],
\label{pss52}
\end{eqnarray}
where
\begin{eqnarray}
\varepsilon_{0}(x)=(2x)^{-3}\left\{3x(1+x^{2})^{1/2}(1+2x^{2})-8x^{3}-3\ln{\big[x+(1+x^{2})^{1/2}\big]}\right\}
\label{pss53}
\end{eqnarray}
is the contribution of an ideal system per one electron, $-3\alpha_{0}x(4\piup)^{-1}$ is the contribution of interactions of the Hartree-Fock approximation, $\alpha_{0}^{2}\varepsilon_{\text c}(x)$  is the correlation contribution. According to our calculations, the $\varepsilon_{\text c}(x)$ can be approximated with the following expression:
\begin{align}
&\varepsilon_{\text c}(x)=-\frac{b_{0}}{2}\int\limits_{0}^{x}\rd t\frac{b_{1}a+t^{1/2}}{t^{3/2}+tb_{1}a+b_{2}t^{1/2}a^{2}+b_{3}a^{3}}
\frac{1+a_{1}t+a_{2}t^{2}}{1+td_{0}}\,,\nonumber\\
&a=(\alpha_{0}\eta)^{1/2};\quad a_{1}=2.25328;\quad a_{2}=4.87991;\quad d_{0}=0.924022;\nonumber\\
&b_{0}=0.0621814;\quad b_{1}=9.81379;\quad b_{2}=2.82214;\quad b_{3}=0.73701.
\label{pss54}
\end{align}
At $a_{1}=a_{2}=d_{0}=0$, the expression matches the approximation for the correlation energy \cite{ref25}, which is calculated using the Monte-Carlo method  \cite{ref26} $\varepsilon_{\text c}^{\text{MC}}(x)$. As we can see in figure~\ref{figs07pss}, in the range $x\leqslant1$, the expression \eqref{pss54} is close to $\varepsilon_{\text c}^{\text{MC}}(x)$, and the deviation $\varepsilon_{\text c}(x)$ from $\varepsilon_{\text c}^{\text{MC}}(x)$ in the region $x>1$ is caused by different asymptotics of these functions $\varepsilon_{\text c}(x)\rightarrow-\frac{b_{0}}{2}\frac{a_{2}}{d_{0}}x+\ldots,$ $\varepsilon_{\text c}^{\text{MC}}(x)\rightarrow-\frac{b_{0}}{2}\ln{x}+\ldots$ at $x\gg1$.

In order to calculate the contributions of electron-nuclear interactions in the products of structure factors in the formula~\eqref{pss46}, we have selected one-particle and two-particle sums by the coordinates of nuclei, and the three-nuclear effective interactions were neglected. In this approach,
\begin{eqnarray}
E\simeq E_{\text e}+E_{\text{pol}}+E_{\text{conf}}\,,
\label{pss55}
\end{eqnarray}
where $E_{\text{pol}}$ is the polarization energy of electron fluid by nuclei, which does not depend on the structure of the nuclear subsystem,
\begin{align}
&E_{\text{pol}}=E^{(2)}_{\text{pol}}+E^{(3)}_{\text{pol}}\,,\nonumber\\
&E^{(2)}_{\text{pol}}=-\frac{z_{1}^{2}N_{1}+z_{2}^{2}N_{2}}{2!V^{2}}
\sum\limits_{\mathbf{q}}V_{q}^{2}\,\tilde{\mu}_{2}(\mathbf{q},-\mathbf{q}|0),\nonumber\\
&E^{(3)}_{\text{pol}}=-\frac{z_{1}^{3}N_{1}+z_{2}^{3}N_{2}}{3!V^{3}}\sum\limits_{\mathbf{q}_{1},\mathbf{q}_{2}}
V_{q_{1}}V_{q_{2}}V_{-q_{1}-q_{2}}\tilde{\mu}_{3}(\mathbf{q}_{1},\mathbf{q}_{2},-\mathbf{q}_{1}-\mathbf{q}_{2}|0).
\label{pss56}
\end{align}
The configuration energy is determined by the structure of a nuclear subsystem, and expressed through an effective two-particle potential of interactions of the nuclei, which is formed by two- and three-electron correlations:
\begin{align}
&E_{\text{conf}}=\sum\limits_{i=1,2}\frac{z_{i}^{2}}{2!V}\sum\limits_{\mathbf{q}}V_{q}\bigg[1-\frac{V_{q}}{V}
\tilde{\mu}_{2}(\mathbf{q},-\mathbf{q}|0)
-\frac{z_{i}}{V^{2}}\sum\limits_{\mathbf{q}_{1}}V_{q_{1}}V_{-q-q_{1}}
\tilde{\mu}_{3}(\mathbf{q},\mathbf{q}_{1},-\mathbf{q}-\mathbf{q}_{1}|0)\bigg]S_{2}^{(i)}(\mathbf{q},-\mathbf{q})\nonumber\\
&+\frac{z_{1}z_{2}}{V}\sum\limits_{\mathbf{q}}V_{q}\biggl[1-\frac{V_{q}}{V}\,
\tilde{\mu}_{2}(\mathbf{q},-\mathbf{q}|0)-\frac{z_{1}+z_{2}}{2V^{2}}
\sum\limits_{\mathbf{q}_{1}}V_{q_{1}}V_{-q-q_{1}}\,
\tilde{\mu}_{3}(\mathbf{q},\mathbf{q}_{1},-\mathbf{q}-\mathbf{q}_{1}|0)\biggr]S^{(1)}_{\mathbf{q}}S^{(2)}_{-\mathbf{q}}\,,\nonumber\\
&S_{2}^{(i)}(\mathbf{q},-\mathbf{q})=\sum\limits_{j_{1}\ne j_{2}=1}^{N_{i}}\exp\big[\ri(\mathbf{q},\mathbf{R}_{j_{1}}-\mathbf{R}_{j_{2}})\big].
\label{pss57}
\end{align}

\subsection{The approximation of two-electron correlations}

Let us rewrite the component $E^{(2)}_{\text{pol}}$,  calculated in the local field approximation, in the form
\begin{eqnarray}
E^{(2)}_{\text{pol}}=N_{\text e}m_{0}c^{2}\frac{\langle z^{2}\rangle}{\langle z\rangle}\alpha_{0}^{3/2}\varepsilon^{(2)}_{\text{pol}}(x),
\label{pss58}
\end{eqnarray}
 where the dimensionless function $\varepsilon^{(2)}_{\text{pol}}(x)$ is of the same order as  $\varepsilon_{\text c}(x)$, and $\langle z^{n}\rangle=(z_{1}^{n}N_{1}+z_{2}^{n}N_{2})(N_{1}+N_{2})^{-1}$. The function $\varepsilon^{(2)}_{\text{pol}}(x)$ can be approximated by the following expression:
\begin{align}
&\varepsilon_{\text{pol}}^{(2)}(x)=-\int\limits_{0}^{x}\frac{c_{0}+c_{1}t+c_{2}t^{2}+c_{3}t^{3}}{1+d_{1}t+d_{2}t^{2}+d_{3}t^{3}}\,\rd t;\nonumber\\
&c_{0}=4.06151;\quad c_{1}=32.6118;\quad c_{2}=-43.6587;\quad c_{3}=104.13;\nonumber\\
&d_{1}=73.8252;\quad d_{2}=-67.1028;\quad d_{3}=189.781.
\label{pss59}
\end{align}
As it is shown in figure~\ref{figs07pss}, $\varepsilon^{(2)}_{\text{pol}}(x)$ has linear asymptotics at $x\gg1$, as well as $\varepsilon_{\text c}(x)$. However, the polarization energy $E^{(2)}_{\text{pol}}$ exceeds the correlation energy of the reference system by about $\langle z\rangle\alpha_{0}^{-1/2}\approx10\langle z\rangle$ times, and for $\langle z\rangle\sim10$ it is comparable with $E_{\text{HF}}$.

The configuration energy was calculated in the coordinate representation and, introducing the effective two-nuclear interactions, yielded
\begin{align}
&V_{2}^{i_{1},i_{2}}\big(\mathbf{R}_{j_{1}}^{(i_{1})}-\mathbf{R}_{j_{2}}^{(i_{2})}\big)=\frac{1}{V}\sum\limits_{\mathbf{q}}V_{2}(q)
\exp\left\{\ri\big[\mathbf{q},\mathbf{R}_{j_{1}}^{(i_{1})}-\mathbf{R}_{j_{2}}^{(i_{2})}\big]\right\},\nonumber\\
&V_{2}(q)=V_{q}\left[1-\frac{V_{q}}{V}\tilde{\mu}_{2}(\mathbf{q},-\mathbf{q}|0)\right].
\label{pss60}
\end{align}
In the formula~\eqref{pss60} the sum of the vector $\mathbf{q}$ includes a component with $\mathbf{q}=0$. Therefore,
\begin{align}
E^{(2)}_{\text{conf}}&=\frac{1}{V}\sum\limits_{\mathbf{q}\ne0}V_{2}(q)
\left[\frac{1}{2}\sum\limits_{i=1,2}S_{2}^{(i)}(\mathbf{q},-\mathbf{q})z_{i}^{2}+
z_{1}z_{2}S^{(1)}_{\mathbf{q}}S^{(2)}_{-\mathbf{q}}\right]\nonumber\\
&=\frac{1}{2}\sum\limits_{i=1,2}z_{i}^{2}\sum\limits_{j_{1}\ne j_{2}=1}V_{2}\big(\mathbf{R}_{j_{1}}^{(i)}-\mathbf{R}_{j_{2}}^{(i)}\big)+
z_{1}z_{2}\sum\limits_{j_{1}=1}^{N_{1}}\sum\limits_{j_{2}=1}^{N_{2}}V_{2}\big(\mathbf{R}_{j_{1}}^{(1)}-\mathbf{R}_{j_{2}}^{(2)}\big)
-\frac{1}{2}N_{\text e}^{2}\lim\limits_{\mathbf{q}\to0}\biggl[\frac{V_{2}(q)}{V}\biggr].
\label{pss61}
\end{align}
To simplify the calculation of the lattice sum, we adopt a simple model of nuclei distribution:
\begin{eqnarray}
\mathcal{N}_{j}^{(1)}=\frac{N_{1}}{N_{1}+N_{2}}\mathcal{N}_{j}\,,\qquad
\mathcal{N}_{j}^{(2)}=\frac{N_{2}}{N_{1}+N_{2}}\mathcal{N}_{j}\,,
\label{pss62}
\end{eqnarray}
where $\mathcal{N}_{j}$ is the number of all knots on the $j$-th coordination sphere, and $\mathcal{N}_{j}^{(i)}$ is the number of the knots occupied by nuclei with charge $z_{i}$ $(i=1,2)$. In this model,
\begin{eqnarray}
E^{(2)}_{\text{conf}}=\frac{N_{\text e}}{2}\langle z\rangle\sum\limits_{j\geqslant 1}\mathcal{N}_{j}V_{2}(R_{j})
-\frac{N_{\text e}m_{0}c^{2}}{6}
\left[\frac{x^{2}}{(1+x^{2})^{1/2}}-\frac{x\alpha_{0}}{\piup}\right],
\label{pss63}
\end{eqnarray}
where $R_{j}$ is the radius of the $j$-th coordination sphere.

\begin{figure}[!t]
\center{\includegraphics[width=0.7\textwidth]{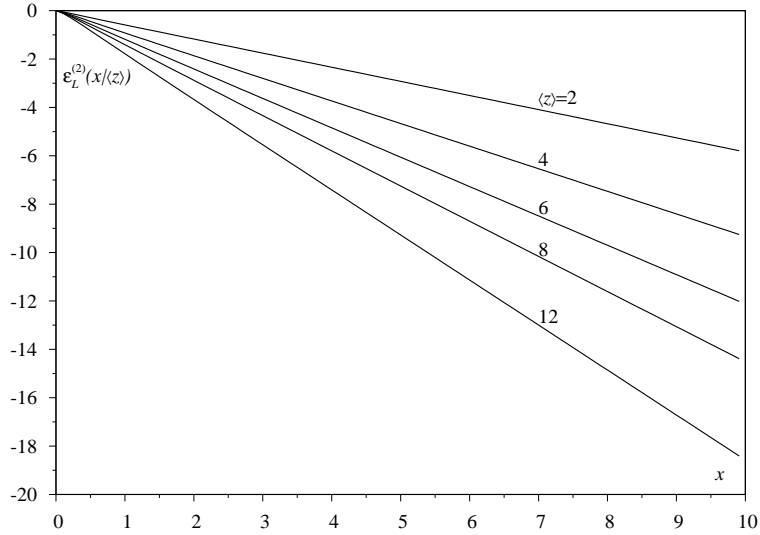}}
\caption{Dependence of the lattice energy $\varepsilon_{\text L}^{(2)}(x|\langle z\rangle)$ on the nuclear charge and the relativistic parameter.}
\label{figs08pss}
\end{figure}

The effective two-particle potential is screened, and at small and medium distances between nuclei it is close to the expression
\begin{eqnarray}
V(R)=\frac{e^{2}}{R}\exp(-R/R_{0}),
\label{pss64}
\end{eqnarray}
and the screening radius
\begin{eqnarray}
R_{0}=\frac{\sqrt{\piup}}{2}\alpha_{0}^{1/2}a_{\text B}\big[x^{1/2}(1+x^{2})^{1/4}\big]^{-1}
\label{pss65}
\end{eqnarray}
is of the order $0.1a_{\text B}$ (where $a_{\text B}=\hbar^{2}/m_{0}e^{2}$ is the Bohr radius). At  large distances, $V_{2}(R)$ oscillates, but with a small amplitude,
\begin{eqnarray}
V_{2}(R)\approx\frac{e^{2}}{a_{\text B}}\left(\frac{R_{0}}{2xR}\right)^{3}\cos(2xR/R_{0}).
\label{pss66}
\end{eqnarray}
The configuration energy for a simple cubic lattice of nuclei is calculated numerically and can be represented as
\begin{eqnarray}
E^{(2)}_{\text{conf}}=N_{\text e}m_{0}c^{2}\langle z\rangle^{2/3}\alpha_{0}\varepsilon_{\text L}^{(2)}(x|\langle z\rangle),
\label{pss67}
\end{eqnarray}
with a dimensionless factor approximated by the expression
\begin{eqnarray}
\varepsilon_{\text L}^{(2)}(x|\langle z\rangle)=-\int\limits_{0}^{x}\frac{a_{1}+ta_{2}+t^{2}a_{3}}{1+ta_{4}+t^{2}a_{5}+t^{3}a_{6}}t\rd t,
\label{pss68}
\end{eqnarray}
where all the coefficients $a_{1},\ldots, a_{6}$ are the functions of $\langle z\rangle$, that is
\begin{eqnarray}
a_{i}(\langle z\rangle)=\frac{a_{i_0}+\langle z\rangle a_{i_1}+\langle z\rangle^{2} a_{i_2}}
{a_{i_3}+\langle z\rangle a_{i_4}+\langle z\rangle^{2} a_{i_5}}.
\label{pss69}
\end{eqnarray}
\begin{table}[!t]
\caption{\label{tab01} Coefficients $a_{ij}$ from \eqref{pss69}.}
\center
\begin{tabular}{|*{7}{c|}}
\hline\hline
$i\bigm\backslash j$ & $0$ & $1$ & $2$ & $3$ & $4$ & $5$\\
\hline\hline
$0$ & $-128.112$ & $-138.098$ & $-3.30915$ & $0$ & $3.74936$ & $0.882489$\\
\hline
$1$ & $-633.899$ & $297.304$ & $-19.5138$ & $1$ &$-0.707632$ & $1.01638$\\
\hline
$2$ & $-1691$ & $216.967$ & $-8.71667$ & $0$ & $1.48694$ & $0.12998$\\
\hline
$3$ & $2.37539$ & $1.74513$ & $0.0417739$ & $0$ & $0.0212583$ & $0.0056168$\\
\hline
$4$ & $913.016$ & $-452.217$ & $30.3618$ & $1$ & $-0.750844$ & $0.702212$\\
\hline
$5$ & $5.68901$ & $-0.704184$ & $0.0277872$ & $0$ & $0.00203554$ & $0.000234399$\\
\hline\hline
\end{tabular}
\end{table}
The cofficients $a_{ij}$ of the formula~(\ref{pss69})  are listed in table~\ref{tab01}.
In figure~\ref{figs08pss} there is shown a dependence of the configuration energy on the relativistic parameter.

\subsection{The effect of three-electron correlations}

According to the formula~\eqref{pss56}, the contribution of three-particle correlations is represented as
\begin{eqnarray}
E^{(3)}_{\text{pol}}=N_{\text e}m_{0}c^{2}\frac{\langle z^{3}\rangle}{\langle z\rangle}\alpha_{0}^{5/2}\varepsilon_{\text{pol}}^{(3)}(x).
\label{pss70}
\end{eqnarray}
The result of numerical calculations is illustrated in figure~\ref{figs07pss}, which shows that for $x>1$ the ratio $\varepsilon_{\text{pol}}^{(3)}(x)\approx0{.}1\varepsilon_{\text{pol}}^{(2)}(x)$ is satisfied. Therefore, at sufficiently large values of nuclei charges, $E^{(3)}_{\text{pol}}$ is not less than the correlation energy of the reference system: at $\langle z\rangle\geqslant 6$, the contribution  $E^{(3)}_{\text{pol}}$ is close to the correlation energy, at $\langle z\rangle\geqslant 12$, it exceeds the correlation energy by 5 times, and at $\langle z\rangle=26$ --- more than by 20~times. The result of numerical calculation $\varepsilon^{(3)}_{\text{pol}}(x)$ is approximated by the expression
\begin{align}
&\varepsilon^{(3)}_{\text{pol}}(x)=-ax-c_{0}\int\limits_{x}^{\infty}\frac{1+c_{1}/t+c_{2}t}{1+d_{1}t+d_{2}t^{2}+d_{3}t^{3}}\,\rd t,\nonumber\\
&a=0.0450;\quad c_{0}=0.12607;\quad c_{1}=-0.93695;\quad c_{2}=78.8552;\nonumber\\
&d_{1}=-23.2602;\quad d_{2}=114.5030;\quad d_{3}=164.060.
\label{pss71}
\end{align}
From the formulae~\eqref{pss58}, \eqref{pss70} it follows that $E^{(3)}_{\text{pol}}/E^{(2)}_{\text{pol}}\sim0.1z\alpha_{0}$, it determines the order of three-electron correlations contribution value.

\begin{figure}[!t]
\center{\includegraphics[width=0.7\textwidth]{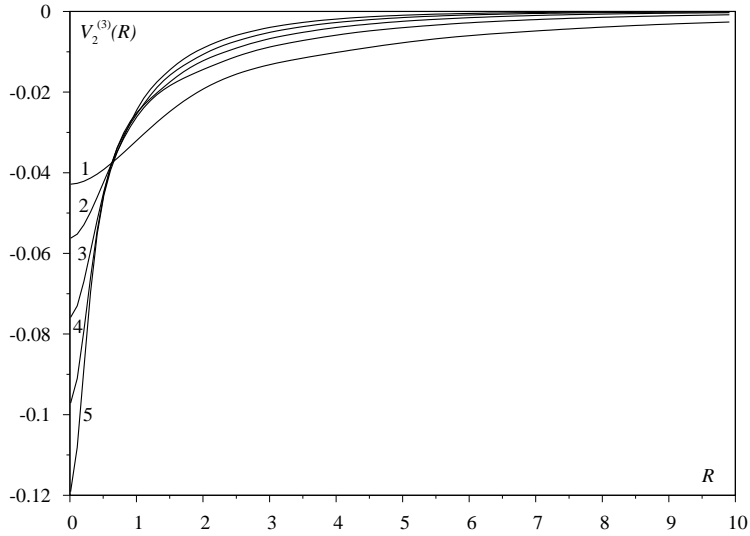}}
\caption{The effective potential of interactions $V_{2}^{(3)}(R)$
at different values of the relativistic parameter (curve~1 ---~$x=1.0$, 2~---~$x=2.0$, 3~---~$x=3.0$, 4~---~$x=4.0$, 5~---~$x=5.0$).}
\label{figs09pss}
\end{figure}

Similar to the formula~\eqref{pss60}, we have introduced a correction to the effective two-nuclear potential by the three-electron correlations
\begin{eqnarray}
V_{2}^{(3)}(R)=-V^{3}\sum\limits_{\mathbf{q}}V_{q}\sum\limits_{\mathbf{q_{1}}}V_{q_{1}}V_{-q-q_{1}}
\tilde{\mu}_{3}(\mathbf{q},\mathbf{q}_{1},-\mathbf{q}-\mathbf{q}_{1}|0)\exp[\ri(\mathbf{q,R})],
\label{pss72}
\end{eqnarray}
where the sum over the vector $\mathbf{q}$ includes the component $\mathbf{q}=0$. As shown in figure~\ref{figs09pss}, $V_{2}^{(3)}(R)$ is a weak attracting potential of the type of quantum package screening potential, which is close to the expression
\begin{eqnarray}
V_{2}^{(3)}(R)=-\frac{e^{2}}{R}\alpha_{0}^{2}
A(x)\left\{1-\exp\left[-\frac{R}{R_{0}}\gamma(x)\right]\right\}\exp(-R/R_{0}).
\label{pss73}
\end{eqnarray}
The functions $A(x)$ and $\gamma(x)$ are determined as follows:
\begin{align}
&A(x)=\frac{8\,I_{2}(x)}{(\piup x)^{2}}\,,\qquad \gamma(x)=\frac{\alpha_{0}^{3/2}x^{1/2}(1+x^{2})^{3/4}}{2\sqrt{\piup}}
\cdot\frac{I_{1}(x)}{I_2(x)};\nonumber\\
&I_{1}(x)=\int\limits_{0}^{\infty}\frac{\rd q_{1}}{\varepsilon(q_{1})}\int\limits_{0}^{\infty}\frac{\rd q_{2}}{\varepsilon(q_{2})}
\int\limits_{-1}^{1}\frac{\rd t}{q_{3}^{2}\varepsilon(q_{3})}\,f_{3}(q_{1},q_{2},t),\qquad
I_{2}(x)=\int\limits_{0}^{\infty}\frac{\rd q}{q^{2}\varepsilon^{2}(q)}J_{3}(q);\nonumber\\
&J_{3}(q)=\frac{(m_{0}c^{2}x^{2})^{2}}{3N_{\text e}}\tilde{\mu}_{3}^{0}(\mathbf{q},-\mathbf{q},0|0);\qquad
f_{3}(q_{1},q_{2},t)=\frac{(m_{0}c^{2}x^{2})^{2}}{3N_{\text e}(1+x^{2})}
\tilde{\mu}_{3}^{0}(\mathbf{q}_{1},\mathbf{q}_{2},-\mathbf{q}_{1}-\mathbf{q}_{1}|0);\nonumber\\
&q_{3}\equiv|\mathbf{q}_{1}+\mathbf{q}_{2}|.
\label{pss74}
\end{align}

\begin{figure}[!t]
\center{\includegraphics[width=0.7\textwidth]{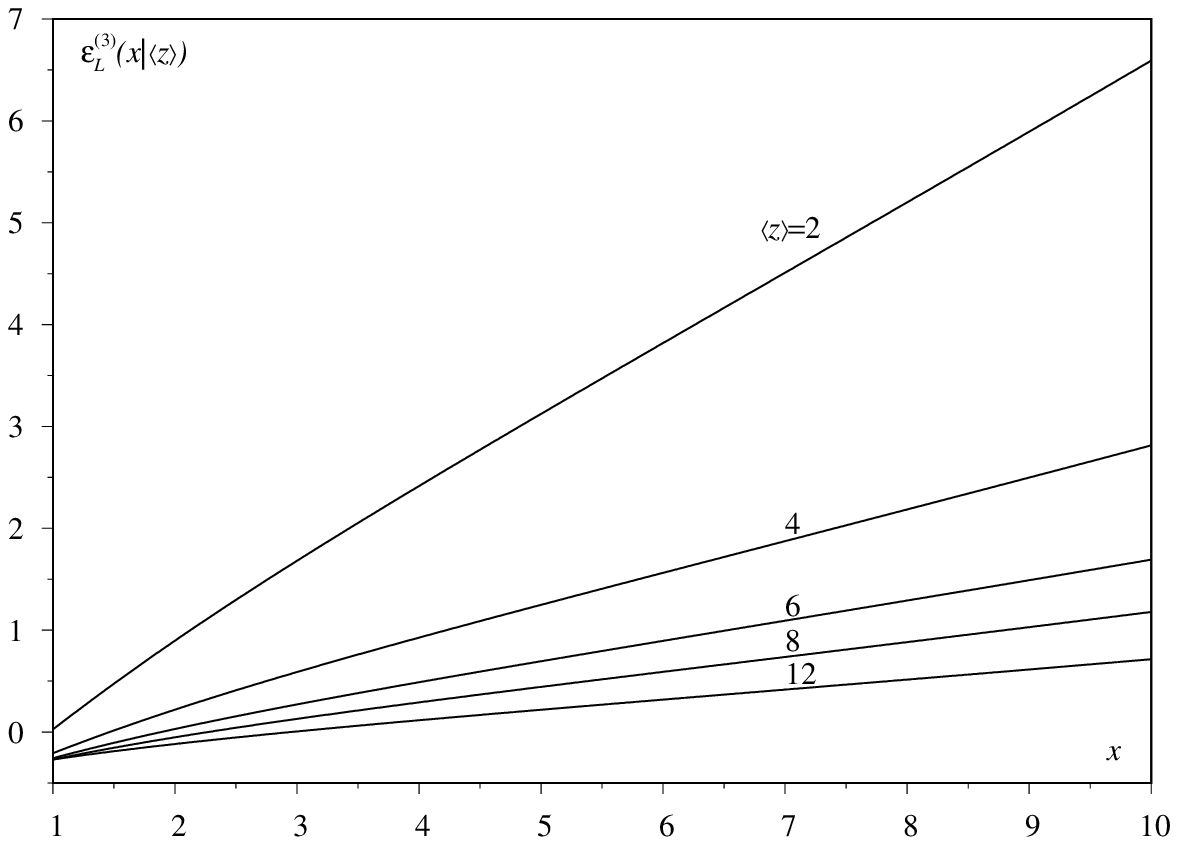}}
\caption{Dependence of the lattice energy $\varepsilon_{\text L}^{(3)}(x|\langle z\rangle)$ on the nuclear charge and the relativistic parameter.}
\label{figs10pss}
\end{figure}

The contribution to the configuration energy of the model by the three-particle correlations in the model~\eqref{pss62} takes the form, similar to the formula~\eqref{pss63}:
\begin{eqnarray}
E^{(3)}_{\text{conf}}=\frac{1}{2}N_{\text e}\langle z^{2}\rangle\sum\limits_{j\geqslant 1}\mathcal{N}_{j}V^{(3)}_{2}(R_{j})+
\frac{4}{3\piup^{2}}N_{\text e}\alpha_{0}^{2}\langle z\rangle m_{0}c^{2}(1+x^{2})^{-1/2}I_{2}(x).
\label{pss75}
\end{eqnarray}
This contribution is calculated for a simple cubic lattice of nuclei and is represented by
\begin{eqnarray}
E^{(3)}_{\text{conf}}=N_{\text e}m_{0}c^{2}\alpha_{0}^{2}\langle z^{2}\rangle\varepsilon_{\text L}^{(3)}(x|\langle z\rangle).
\label{pss76}
\end{eqnarray}
The dependence of a dimensionless factor $\varepsilon_{\text L}^{(3)}(x|\langle z\rangle)$ is illustrated in figure~\ref{figs10pss}. At sufficiently large nuclei charges $\langle z\rangle$ and $x\geqslant 2$, the function $\varepsilon_{\text L}^{(3)}(x|\langle z\rangle)\sim0.1\varepsilon_{\text L}^{(2)}(x|\langle z\rangle)$, but it has a positive sign. It is approximated by the expression
\begin{eqnarray}
\varepsilon_{\text L}^{(3)}(x|\langle z\rangle)=-\frac{a_{0}(\langle z\rangle)+a_{1}(\langle z\rangle)x+a_{2}(\langle z\rangle)x^{2}}{x}\,.
\label{pss77}
\end{eqnarray}
The coefficients of the formula~(\ref{pss77}) are shown in table~\ref{tab02}.
\begin{table}[h!]
\center
\caption{\label{tab02} Coefficients from \eqref{pss77}.}
\vspace{2ex}
\begin{tabular}{|c|c|c|c|}
\hline\hline
$\langle z\rangle$ & $a_{0}$ & $a_{1}$ & $a_{2}$\\
\hline
\hline
$2$ &  $0.160403$ &	$0.422066$ & $-0.707095$\\
\hline
$4$ & $0.118181$ &	$0.351743$ & $-0.320001$\\
\hline
$6$ & $0.0978784$ &	$0.321462$ & $-0.203819$\\
\hline
$8$ & $0.08553532$ &	$0.302602$ & $-0.150039$\\
\hline
$12$ & $0.0701315$ &	$0.278365$ & $-0.100583$\\
\hline
$26$ & $0.04788678$ & $0.236761$ & $-0.0549487$\\
\hline\hline
\end{tabular}
\end{table}

\section{The equation of state of the model at $T=0$~K}

\begin{table}[!t]
{\caption{Dependence of the functions $\mathcal{F}(x)\cdot x^{-4}$, $f_{2}(x)\cdot x^{-4}$ and $f_{3}(x)\cdot x^{-4}$ on the relativistic parameter~$x$ according to the formulae~\eqref{pss79}--\eqref{pss81} and \eqref{pss82}.}
\label{tab2}}
\vspace{2ex}
\begin{center}
\scriptsize
\scalebox{.9}{
\begin{tabular}{|*{10}{c|}}
\hline\hline
\multirow{2}{*}{$x$} & \multirow{2}{*}{$\mathcal{F}(x)\cdot x^{-4}$} & \multicolumn{4}{c|}{$f_{2}(x)\cdot x^{-4\strut}$} & \multicolumn{4}{c|}{$f_{3}(x)\cdot x^{-4}$}\\
\cline{3-10} && $z=2$ & $z=6$ & $z_{1}=7;\, z_{2}=8$ & $z=12$ & $z=2$ & $z=6$ & $z_{1}=7;\,z_{2}=8$ & $z=12$ \\
\hline\hline
$0.5$ &	$0.737488$ &	$-0.025120$ &	$-0.0899288$ &	$-0.113384$ &	$-0.208687$ &	$0.00151018$ &	$0.00503624$ &	 $0.00724728$ &	 $0.0123084$\\
\hline
$0.6$ &	$0.857456$ &	$-0.0249008$ &	$-0.0898362$ &	$-0.113389$ &	$-0.209377$ &	$0.00144201$ &	$0.0045685$ &	 $0.00653182$ &	 $0.0107583$\\
\hline
$0.7$ &	$0.966234$ &	$-0.0247355$ &	$-0.0896437$ &	$-0.11323$ &	$-0.209566$ & $0.00140653$ &	$0.00431288$ &	 $0.00610532$ &	 $0.00988929$\\
\hline
$0.8$ &	$1.06412$ &	$-0.0246194$ &	$-0.0894304$ &	$-0.113012$ &	$-0.209487$ &	$0.00138662$ &	$0.00415774$ &	 $0.00582077$ &	 $0.00935182$\\
\hline
$0.9$ &	$1.15175$ &	$-0.0245407$ &	$-0.0892304$ &	$-0.112788$ &	$-0.20927$ & $0.00137469$ &	$0.00405367$ &	 $0.00561505$ &	$0.00899025$ \\
\hline
$1.0$ &	$1.22991$ &	$-0.0244886$ &	$-0.0890553$ &	$-0.11258$ &	$-0.208989$ &	$0.00136685$ &	$0.00397618$ &	 $0.00545736$ &	 $0.00872662$\\
\hline
$1.1$ &	$1.29949$ &	$-0.0244546$ &	$-0.0889067$ &	$-0.112394$ &	$-0.208686$ &	$0.00136085$ &	$0.00391243$ &	 $0.00533111$ &	 $0.00851882$\\
\hline
$1.2$ &	$1.36139$ &	$-0.0244329$ &	$-0.0887823$ &	$-0.112234$ &	$-0.208385$ &	$0.00135532$ &	$0.00385556$ &	 $0.00522669$ &	 $0.00834323$\\
\hline
$1.3$ &	$1.41647$ &	$-0.0244195$ &	$-0.0886788$ &	$-0.112096$ &	$-0.208099$ &	$0.00134944$ &	$0.00380207$ &	 $0.00513813$ &	 $0.00818666$\\
\hline
$1.4$ &	$1.46551$ &	$-0.0244117$ &	$-0.0885926$ &	$-0.111979$ &	$-0.207833$ &	$0.00134276$ &	$0.0037504$ &	 $0.00506157$ &	$0.008042$ \\
\hline
$1.5$ &	$1.50924$ &	$-0.0244077$ &	$-0.0885206$ &	$-0.111878$ &	$-0.207591$ &	$0.00133514$ &	$0.00370008$ &	 $0.00499435$ &	 $0.00790579$\\
\hline
$1.6$ &	$1.5483$ &	$-0.0244062$ &	$-0.0884603$ &	$-0.111793$ &	$-0.207371$ &	$0.00132663$ &	$0.00365119$ &	 $0.00493463$ &	 $0.00777667$\\
\hline
$1.7$ &	$1.58327$ &	$-0.0244065$ & 	$-0.0884095$ &	$-0.111719$ &	$-0.207173$ &	$0.00131746$ &	$0.00360403$ &	 $0.00488104$ &	 $0.00765433$\\
\hline
$1.8$ &	$1.61463$ &	$-0.024408$ &	$-0.0883666$ &	$-0.111656$ &	$-0.206995$ &	$0.00130788$ &	$0.00355898$ &	 $0.00483256$ &	 $0.00753897$\\
\hline
$1.9$ &	$1.64282$ &	$-0.0244103$ &	$-0.08833$ &	$-0.111602$ &	$-0.206836$ &	$0.0012982$ &	$0.00351632$ &	 $0.00478842$ &	 $0.00743084$\\
\hline
$2.0$ &	$1.66822$ &	$-0.0244131$ &	$-0.0882987$ &	$-0.111555$ &	$-0.206693$ &	$0.00128866$ &	$0.00347628$ &	 $0.00474798$ &	 $0.00733015$\\
\hline
$3.0$ &	$1.82417$ &	$-0.0244459$ &	$-0.0881427$ &	$-0.111308$ &	$-0.205863$ &	$0.00122235$ &	$0.00321213$ &	 $0.00447214$ &	 $0.00668381$\\
\hline
$4.0$ &	$1.89283$ &	$-0.0244708$ &	$-0.0880955$ &	$-0.111226$ &	$-0.205545$ &	$0.00119715$ &	$0.0031007$ &	 $0.00431866$ &	 $0.00642247$\\
\hline
$5.0$ &	$1.92833$ &	$-0.024488$ & 	$-0.0880781$ &	$-0.111192$ &	$-0.205404$ &	$0.00118685$ &	$0.00304877$ &	 $0.00422167$ &	 $0.00630395$ \\
\hline\hline
\end{tabular}}
\end{center}
\end{table}

For the well-known dependence of the model energy on the relativistic parameter we calculate the equation of state of cold degenerate matter using  the expression
\begin{eqnarray}
P(x)=\frac{\rd E}{\rd V}=\frac{x^{4}}{N_{\text e}}\left(\frac{m_{0}c}{\hbar}\right)^{3}(3\piup^{2})^{-1}\frac{\rd E}{\rd x}\,.
\label{pss78}
\end{eqnarray}
Within the accepted approximation
\begin{eqnarray}
P(x)=\frac{\piup m_{0}^{4}c^{5}}{3h^{3}}[\mathcal{F}(x)+f_{2}(x)+f_{3}(x)].
\label{pss79}
\end{eqnarray}
Here,
\begin{eqnarray}
\mathcal{F}(x)=x(2x^{2}-3)(1+x^{2})^{1/2}+3\ln{[x+(1+x^{2})^{1/2}]}
\label{pss80}
\end{eqnarray}
is the contribution of the ideal degenerate relativistic spatially homogeneous electron gas;
\begin{eqnarray}
f_{2}(x)=-2\alpha_{0}x^{4}\biggl\{\frac{1}{\piup}-\frac{4}{3}\frac{\rd}{\rd x}\biggl[\langle z\rangle^{2/3}
\varepsilon_{\text L}^{(2)}(x|\langle z\rangle)+
\frac{\langle z^{2}\rangle}{\langle z\rangle}\alpha_{0}^{1/2}\varepsilon^{(2)}_{\text{pol}}(x)+\alpha_{0}\varepsilon_{\text c}(x)\biggr]\biggr\}
\label{pss81}
\end{eqnarray}
is the contribution of Coulomb interactions in the two-electron correlations approximation;
\begin{eqnarray}
f_{3}(x)=8\alpha_{0}^{2}x^{4}\frac{\rd}{\rd x}\biggl[\langle z^{2}\rangle\varepsilon_{\text L}^{(3)}(x|\langle z\rangle)+
\frac{\langle z^{3}\rangle}{\langle z\rangle}\alpha_{0}^{1/2}\varepsilon^{(3)}_{\text{pol}}(x)\biggr]
\label{pss82}
\end{eqnarray}
is the contribution of the three-particle electron correlations.

In the region $x\geqslant 1$, all contributions [with the exception of $\varepsilon_{\text L}^{(3)}(x|\langle z\rangle)$] to the model energy caused by interactions are negative monotonously decreasing functions of the relativistic parameter. In the two-electron correlations approximation, the equation of state \eqref{pss79} numerically is very close to the result of Salpeter \cite{ref06}.

\begin{figure}[!t]
\center{\includegraphics[width=0.7\textwidth]{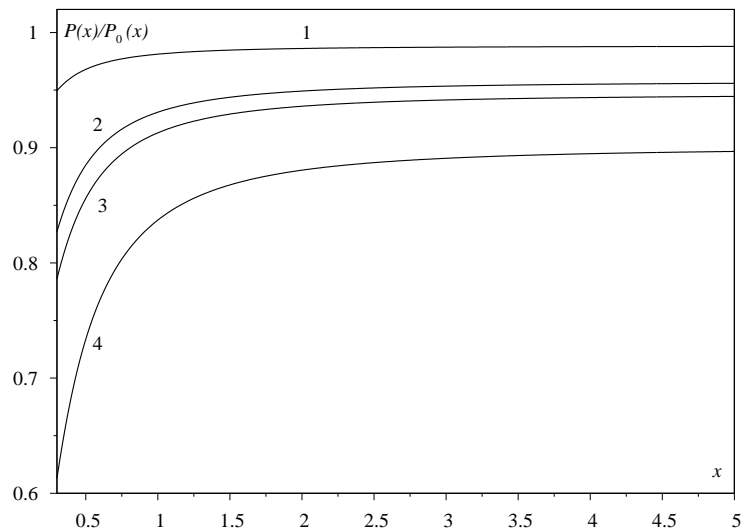}}
\caption{The ratio of pressure $P(x)$ to the pressure of the ideal relativistic electron gas $P_{0}(x)$ as function of the relativistic parameter and nuclear charge (curve~1 ---~$z_{1}=z_{2}=2$, 2~---~$z_{1}=z_{2}=6$, 3~---~$z_{1}=7;\,z_{2}=8;$ 4~---~$z_{1}=z_{2}=12$).}
\label{figs11pss}
\end{figure}

In table~\ref{tab2} there is shown a dependence of  terms $\mathcal{F}(x)$, $f_{2}(x)$, $f_{3}(x)$ on the relativistic parameter for the helium  ($z_{1}=z_{2}=2$), carbon ($z_{1}=z_{2}=6$), nitrogen-oxygen ($z_{1}=7$, $z_{2}=8$; $N_{1}=N_{2}$) and magnesium ($z_{1}=z_{2}=12$) dwarf models. A relative decrease of pressure caused by the interactions $[\mathcal{F}(x)+f_{2}(x)+f_{3}(x)]\mathcal{F}^{-1}(x)$ is illustrated in figure~\ref{figs11pss}.

\section{Conclusions}

Reference system approach was adapted for the description of a degenerate relativistic electron subsystem. There were investigated the features of two- and three-particle correlation functions in a wide domain of a relativistic parameter, as well as there was obtained
an exact expression for the static two-particle correlation function. For the first time, the expressions for the three-particle correlation function were  approximated and  the local field correction of interacting relativistic electron gas was studied, which is the basis for the calculation of the energy and structural characteristics of degenerated dwarfs. The energy of the ground state of the electron-nuclear model, as well as the equation of state of the model, have been calculated in a wide range of the relativistic parameter at absolute zero temperature. As it was shown in our calculation, the contributions of Coulomb interactions to the energy of the ground state and pressure, caused by two-electron correlations are important and increase with an increase of nuclear charge. The contributions caused by  three-electron correlations are much smaller but they exceed the contribution of correlation energy of the electron fluid, especially at large values of the nuclear charge.

\ukrainianpart

\title{Базисний підхід в теорії вироджених карликів}
\author{М.В. Ваврух, Д.В. Дзіковський, Н.Л. Тишко}
\address{
Львівський національний університет імені Івана Франка, кафедра астрофізики, \\вул. Кирила і Мефодія, 8, 79005 Львів, Україна
}

\makeukrtitle

\begin{abstract}
\tolerance=3000%
Базисний підхід нерелятивістської теорії електронної рідини адаптовано до розрахунку характеристик електрон-ядерної моделі при густинах, що відповідають виродженим карликам. Розраховано дво- та тричастинкові кореляційні функції виродженого релятивістського електронного газу в імпульсно-частотному представленні у наближенні локального поля. В адіабатичному наближенні обчислено основні внески кулонівських взаємодій в енергію та рівняння стану моделі при $T=0$~K.
\keywords електрон-ядерна модель, кореляційні функції, поправка на локальне поле, енергія моделі при $T=0$~K, рівняння стану

\end{abstract}


\begin{thebibliography}{26}
\bibitem{ref01} Adams W.S., Publ. Astron. Soc. Pac., 1915, \textbf{27}, 236, \doi{10.1086/122440}.
\bibitem{ref02} Chandrasekhar S., Astrophys. J., 1931, \textbf{74}, 81, \doi{10.1086/143324}.
\bibitem{ref03} Chandrasekhar S., Mon. Not. R. Astron. Soc., 1935, \textbf{95}, 676, \doi{10.1093/mnras/95.8.676}.
\bibitem{ref04} James R.A., Astrophys. J., 1964, \textbf{140}, 552, \doi{10.1086/147949}.
\bibitem{ref05} Zeldovich Ya.B., Novikov I.D., Relativistic Astrophysics, Nauka, Moscow, 1967 (in Russian). 
\bibitem{ref06} Salpeter E.E., Astrophys. J., 1961, \textbf{134}, 669, \doi{10.1086/147194}.
\bibitem{ref07} Vavrukh M.V., Smerechynskyi S.V., Astron. Rep., 2012, \textbf{56}, No.~5, 363, \doi{10.1134/S1063772912050071}.
\bibitem{ref08} Vavrukh M.V., Smerechinskii S.V., Astron. Rep., 2013, \textbf{57}, No.~12, 913, \doi{10.1134/S1063772913100065}.
\bibitem{ref09} Ostriker J.P., Hartwick F.D.A., Astrophys. J., 1968, \textbf{153}, 797, \doi{10.1086/149706}.
\bibitem{ref10} Shapiro S.L., Teukolsky S.A., Black Holes, White Dwarfs and Neutron Stars, Cornell University, Ithaca,
New York, 1983.
\bibitem{ref11} Kaplan S.A., Sci. Notes Ivan Franko Lviv State Univ., Ser. Phys. Math., 1949, \textbf{4}, 109 (in Russian). 
\bibitem{ref12} Hamada T., Salpeter E., Astrophys. J., 1961, \textbf{134}, 683, \doi{10.1086/147195}.
\bibitem{ref13} Tremblay P.-E., Bergeron P.,  Gianninas A., Astrophys. J., 2011, \textbf{730}, 128, \doi{10.1088/0004-637X/730/2/128}.
\bibitem{ref14} DeGennaro S., von Hippel T., Winget D.E., Kepler S.O., Nitta A., Koester D., Althaus L., Astron. J., 2008, \textbf{135}, 1, \doi{10.1088/0004-6256/135/1/1}. 
\bibitem{ref15} Vavrukh M., Tyshko N., Smerechynskyj S., Math. Model. Comput., 2014, \textbf{1}, 264.
\bibitem{ref16} Vavrukh M., Krohmalskii T., Phys. Status Solidi B, 1991, \textbf{168}, 519, \doi{10.1002/pssb.2221680213}.
\bibitem{ref17} Vavrukh M., Krohmalskii T., Phys. Status Solidi B, 1992, \textbf{169}, 451, \doi{10.1002/pssb.2221690218}.
\bibitem{ref19} Lloyd P., Sholl C., J. Phys. C: Solid State Phys., 1968, \textbf{1}, 1620, \doi{10.1088/0022-3719/1/6/319}. 
\bibitem{ref20} Brovman E.G., Kagan Yu., Zh. Eksp. Teor. Fiz., 1973, \textbf{63}, 1937 (in Russian). 
\bibitem{ref18} Gell-Mann M., Brueckner K., Phys. Rev., 1957, \textbf{106}, 364, \doi{10.1103/PhysRev.106.364}.
\bibitem{ref21} Brovman E.G., Kholas A., Zh. Eksp. Teor. Fiz., 1974, \textbf{66}, 1877 (in Russian). 
\bibitem{ref23} Hwa  R.C.,  Teplitz V.L., Homology and Feynman Integrals, W.A. Benjamin, Inc., New York, 1966. 
\bibitem{ref22} Vavrukh M., Vavrukh N., Low Temp. Phys., 1996, \textbf{22}, 767.
\bibitem{ref24} Kohn W., Luttinger J., Phys. Rev., 1960, \textbf{118}, 41, \doi{10.1103/PhysRev.118.41}.
\bibitem{ref25} Vosko S.H., Wilk L., Nusair M., Can. J. Phys., 1980, \textbf{58}, 1200, \doi{10.1139/p80-159}.
\bibitem{ref26} Ceperley D., Alder B., Phys. Rev. Lett., 1980, \textbf{45}, 566, \doi{10.1103/PhysRevLett.45.566}.
\end{thebibliography}
\end{document}